\documentclass[12pt]{article}

\usepackage{amssymb}
\usepackage{amsfonts}
\usepackage{amstext}
\usepackage{amsthm}
\usepackage{axodraw}
\usepackage{bbm}
\usepackage{epsfig}
\usepackage{booktabs}
\usepackage{graphicx}
\usepackage{theorem}
\usepackage{array}
\usepackage[numbers, sort]{natbib}
\usepackage{hyperref}
\usepackage{color}
\usepackage[toc, page]{appendix}
\usepackage{float}
\usepackage{dsfont}

\usepackage{rotating}
\usepackage{a4}
\usepackage{a4wide}

\def\L{\mathcal L}
\def\e{\varepsilon}

\newcommand{\wt}{\widetilde}

\newcommand{\be}{\begin{equation}}
\newcommand{\ee}{\end{equation}}
\newcommand{\beq}{\begin{equation}}
\newcommand{\eeq}{\end{equation}}
\newcommand{\bea}{\begin{eqnarray}}
\newcommand{\eea}{\end{eqnarray}}

\newcommand{\D}{\displaystyle}
\def\ra{\rightarrow}

\newcommand{\mt}{$\mu$-$\tau$~}

\newcommand{\s}{\Sigma}

\makeatletter

\newcommand{\Rmnum}[1]{\expandafter\@slowromancap\romannumeral #1@}
\makeatother

\begin{document}

\def\a{\alpha}
\def\b{\beta}
\def\c{\chi}
\def\d{\delta}
\def\e{\epsilon}
\def\f{\phi}
\def\g{\gamma}
\def\h{\eta}
\def\i{\iota}
\def\j{\psi}
\def\k{\kappa}
\def\l{\lambda}
\def\m{\mu}
\def\n{\nu}
\def\o{\omega}
\def\p{\pi}
\def\q{\theta}
\def\r{\rho}
\def\s{\sigma}
\def\t{\tau}
\def\u{\upsilon}
\def\x{\xi}
\def\z{\zeta}
\def\D{\Delta}
\def\F{\Phi}
\def\G{\Gamma}
\def\J{\Psi}
\def\L{\Lambda}
\def\O{\Omega}
\def\P{\Pi}
\def\Q{\Theta}
\def\S{\Sigma}
\def\U{\Upsilon}
\def\X{\Xi}

%Varletters
\def\ve{\varepsilon}
\def\vf{\varphi}
\def\vr{\varrho}
\def\vs{\varsigma}
\def\vq{\vartheta}

\def\dg{\dagger}                                     % hermitian conjugate
\def\ddg{\ddagger}                                   % double dagger
\def\wt#1{\widetilde{#1}}                    % big tilde
\def\mt{\widetilde{m}_1}
\def\mti{\widetilde{m}_i}
\def\rt{\widetilde{r}_1}
\def\mtt{\widetilde{m}_2}
\def\mttt{\widetilde{m}_3}
\def\rtt{\widetilde{r}_2}
\def\mb{\overline{m}}
\def\VEV#1{\left\langle #1\right\rangle}        % < >
\def\be{\begin{equation}}
\def\ee{\end{equation}}
\def\ds{\displaystyle}
\def\ra{\rightarrow}

\def\bea{\begin{eqnarray}}
\def\eea{\end{eqnarray}}
\def\NO{\nonumber}
\def\Bar#1{\overline{#1}}

% Journal abbreviations (preprints)

\def\pl#1#2#3{Phys.~Lett.~{\bf B {#1}} ({#2}) #3}
\def\np#1#2#3{Nucl.~Phys.~{\bf B {#1}} ({#2}) #3}
\def\prl#1#2#3{Phys.~Rev.~Lett.~{\bf #1} ({#2}) #3}
\def\pr#1#2#3{Phys.~Rev.~{\bf D {#1}} ({#2}) #3}
\def\zp#1#2#3{Z.~Phys.~{\bf C {#1}} ({#2}) #3}
\def\cqg#1#2#3{Class.~and Quantum Grav.~{\bf {#1}} ({#2}) #3}
\def\cmp#1#2#3{Commun.~Math.~Phys.~{\bf {#1}} ({#2}) #3}
\def\jmp#1#2#3{J.~Math.~Phys.~{\bf {#1}} ({#2}) #3}
\def\ap#1#2#3{Ann.~of Phys.~{\bf {#1}} ({#2}) #3}
\def\prep#1#2#3{Phys.~Rep.~{\bf {#1}C} ({#2}) #3}
\def\ptp#1#2#3{Progr.~Theor.~Phys.~{\bf {#1}} ({#2}) #3}
\def\ijmp#1#2#3{Int.~J.~Mod.~Phys.~{\bf A {#1}} ({#2}) #3}
\def\mpl#1#2#3{Mod.~Phys.~Lett.~{\bf A {#1}} ({#2}) #3}
\def\nc#1#2#3{Nuovo Cim.~{\bf {#1}} ({#2}) #3}
\def\ibid#1#2#3{{\it ibid.}~{\bf {#1}} ({#2}) #3}

\newcommand{\stefan}[1]{{\color{blue} #1}} % needs the package 'color'
\newcommand{\pasquale}[1]{{\color{magenta} #1}} % needs the package 'color'
\newcommand{\david}[1]{{\color{red} #1}} % needs the package 'color'
\newcommand{\steve}[1]{{\color{green} #1}} % needs the package 'color'

\title{
%{\normalsize \mbox{ }\hfill %\begin{minipage}{3cm} %MPP-2005-118 %\end{minipage}}\\ \vspace*{15mm}
\bf Leptogenesis in the two right-handed neutrino model revisited}
\author{{\Large S.~Antusch$^{a,b}$, P.~Di~Bari$^c$, D.~A.~Jones$^c$, S.~F.~King$^c$}
\\[4mm]
$^a$ {\it\small Max-Planck-Institut f\"{u}r Physik} {\it\small (Werner-Heisenberg-Institut)} \\[-2mm]
{\it\small F\"{o}hringer Ring 6, 80805 M\"{u}nchen, Germany}\\
$^b$ {\it\small Department of Physics, University of Basel, Klingelbergstr.~82, CH-4056 Basel, Switzerland}\\
$^c$ {\it\small School of Physics and Astronomy},{\it\small University of
Southampton,} {\it\small  Southampton, SO17 1BJ, U.K.}
}
\maketitle \thispagestyle{empty}

%\vspace{-19mm} %\centerline{\date{\today}}

\begin{abstract}
We revisit leptogenesis in the minimal non-supersymmetric type I see-saw mechanism with
two right-handed (RH) neutrinos, including flavour effects and allowing
both RH neutrinos $N_1$ and $N_2$ to contribute, rather than just the lightest RH
neutrino $N_1$ that has hitherto been considered. By performing scans over parameter space in terms of the single
complex angle $z$ of the orthogonal matrix $R$, for a range of PMNS parameters,
we find that in regions around $z\sim \pm \pi /2$,
for the case of a normal mass hierarchy, the
$N_2$ contribution can dominate the contribution to leptogenesis, allowing the lightest RH
neutrino mass to be decreased by about an order of magnitude in these regions,
down to $M_1\sim 1.3 \times 10^{11}\,{\rm GeV}$ for vanishing initial $N_2$-abundance,
with the numerical results supported by analytic estimates.
We show that the regions around $z\sim \pm \pi /2$
correspond to light sequential dominance, so the new results in this paper may be relevant to
unified model building.
\end{abstract}
\newpage

\section{Introduction}

Current low energy neutrino data with bi-large mixing can be minimally explained within
the non-supersymmetric type I seesaw mechanism \cite{seesaw} with only
two RH neutrinos \cite{King:1999mb}. This can be regarded as a limiting case of three RH neutrinos where one
of the RH neutrinos decouples from the see-saw mechanism either because it is very heavy or because its Yukawa
couplings are very weak. With only two RH neutrinos it is straightforward to see that the
lightest left-handed neutrino mass has to vanish. Since the number of parameters is greatly reduced, this model has attracted
attention in connection with the possibility to test it, especially when successful leptogenesis \cite{fy}
is required in addition to the constraints from low energy neutrino data \cite{ibarra1}.
The number of parameters (11) is however still high enough that the parameter space cannot be yet over-constrained by
combining together low energy neutrino experiments and successful leptogenesis.
Due to its minimality, there has been a great deal of attention paid to the
two RH neutrino type I non-supersymmetric see-saw mechanism,
(i) in the unflavoured case, then (ii) including flavour dependent effects, as follows.

(i) In the unflavoured approximation,
in \cite{fgy} it was shown that, imposing double texture zeros in the neutrino Dirac mass matrix, it is possible to make connections between the sign of the baryon asymmetry of the Universe and the sign of $C\!P$ violation in neutrino mixing. Similar results also apply when the two RH neutrino model is regarded as a limiting case of three RH neutrinos when the heaviest RH neutrino of mass $M_3 \gg 10^{14}\,{\rm GeV}$ decouples from the see-saw mechanism \cite{King:2002qh,ct}.
In \cite{ir} a systematic leptogenesis study of this model was performed, also
with texture zeros in the neutrino Dirac mass matrix, considering lepton flavour
violating processes within supersymmetric scenarios.
 Leptogenesis with two RH neutrinos has been also studied beyond the hierarchical limit,
obtaining the precise conditions both for the degenerate limit and for the hierarchical limit to be recovered \cite{beyond}.

(ii) A first analysis that included flavour effects \cite{flavoreffects} was presented in \cite{abada2}. The close connection between the CP violation for leptogenesis and the observable leptonic Dirac CP phase in models with two texture zeros and in the limit of a nearly decoupled RH neutrino, which corresponds to the two RH neutrino case, has been studied in
\cite{Antusch:2006cw}. In \cite{mp} a flavoured analysis of leptogenesis within the two RH neutrino model has shown that for an
inverted hierarchical spectrum and if the condition $-\sin\theta_{13}\cos\delta \gtrsim 0.15 $ applies, then the model
can work only in presence of CP violating Majorana and Dirac phases.
In \cite{dirac} the two RH neutrino limit was considered within a leptogenesis scenario when  $C\!P$ violation is
uniquely stemming from the Dirac phase.
The leptogenesis lower bound on $M_1$ in the two RH neutrino model has been
studied in \cite{bounds}, showing that in the absence of PMNS phases and for inverted hierarchy
the bound is much more stringent, confirming the importance of the PMNS phases for this case,
as also pointed out in \cite{mp}.

In this paper we go beyond the above analyses of
leptogenesis in the minimal non-supersymmetric type I see-saw mechanism with two RH neutrinos,
allowing both RH neutrinos $N_1$ and $N_2$ to contribute, rather than just the lightest RH
neutrino $N_1$ that has hitherto been considered 
\footnote{
So far, no dedicated study of leptogenesis in the 2 RH neutrino model exists, where the $N_2$ contribution has been taken into account.
For works where the importance of the $N_2$ contribution was emphasized (in general scenarios) see \cite{N2dominated,N2importance,bounds,us}.
}.
We also emphasize the role of a self-energy contribution to the flavoured $C\!P$ asymmetries which is
frequently ignored for $N_1$ leptogenesis in the hierarchical limit,
but which will prove to be crucial for successful $N_2$ leptogenesis in the hierarchical limit
assumed in this paper.
By performing scans over parameter space in terms of the single
complex angle $z$ of the orthogonal matrix $R$, for a range of PMNS parameters,
we find that in regions around $z\sim \pm \pi /2$,
for the case of a normal mass hierarchy, the
$N_2$ contribution can dominate the contribution to leptogenesis, allowing the lightest RH
neutrino mass to be decreased by about an order of magnitude in these regions,
down to $M_1\sim 1.3\times 10^{11}\,{\rm GeV}$ in the case of initial  vanishing 
$N_2$-abundance. Interestingly, these regions
correspond to so-called light sequential dominance, in which
$N_1$ dominantly contributes to the atmospheric neutrino mass $m_3$, while
$N_2$ dominantly contributes to the solar neutrino mass $m_2$.

In Section 2 we  set up the general notation for leptogenesis in the two RH neutrino model.
In section 3 we solve the Boltzmann equations finding an analytical solution for the final
asymmetry. In section 4 we express the $C\!P$ asymmetries within the orthogonal parametrization.
In Section 5 we show the allowed regions contrasting the $N_1$ production and the $N_2$ production
and showing that new regions appear thanks to the $N_2$ contribution to the final asymmetry.
In Section 6 we show that these regions correspond to light sequential dominance.
In Section 7 we draw the conclusions.

%------------------------------------
\section{General Set up and notation}
%------------------------------------

The Yukawa part of the Lagrangian in a SM extension to include heavy RH neutrinos is given by,
\bea
-{\cal L}_Y= Y_e \overline{L}H l_R+ Y_{\nu}\overline{L}\tilde{H} N_R +\frac{1}{2} \overline{{N}^c_R} M N_R+\rm{h.c.} \, ,
\eea
where $L$ and $H$ are the left-handed lepton doublet and Higgs doublet respectively, $l_R$ the RH charged
singlet and $N_R$ the RH neutral singlet. $Y_e$ and $Y_\nu$ are the Yukawa couplings and $M$ the RH
Majorana neutrino mass matrix. In the above equation $\tilde{H} = -i\sigma_2H^*$. After electroweak symmetry breaking we
get the Dirac mass matrix $m_D=Y_\nu v$, where $v$ is the vacuum expectation value of the Higgs doublet. If we consider
$n$ generations of heavy RH neutrinos $N_R$, then the Dirac mass matrix $m_D$ is a $3 \times n$ matrix and the Majorana mass matrix $M$ is a $n \times n$ matrix. The $(3+n) \times (3+n)$  neutrino mass matrix turns out to be,
\bea
-{\cal L}_m=\pmatrix { \bar{\nu}_L & \bar{N}^c_R} \pmatrix{ 0 & m_D \cr   m_D^T & M}
  \pmatrix{ \nu^c_L \cr N_R}+\rm{h.c}.  \;\; .
\eea

In the see-saw limit, $M\gg m_D$, the spectrum of neutrino mass eigenstates
splits in two sets: $n$ very heavy neutrinos, $N_1,\dots,N_n$
respectively with masses $M_1\leq M_2 \leq  \dots \leq M_n$ and almost coinciding with
the eigenvalues of $M$, and 3 light neutrinos with masses $m_1\leq m_2\leq m_3$ for normal hierarchy (NH) and $m_3\leq m_1\leq m_2$ for inverted hierarchy (IH),
the eigenvalues of the light neutrino mass matrix.

Once the $n$ heavy RH neutrino fields get integrated out from the theory, one obtains the $3 \times 3$
light neutrino mass matrix, up to an irrelevant overall sign,  as
\bea
m_{\nu}\simeq m_D\, M^{-1}\, m_D^T, \label{eq:typeI}
\eea
where we neglected terms higher than ${\cal O}(M^{-2})$. The heavy neutrino mass matrix is approximately given
by $M$. %, where we neglect %terms higher than ${\cal O}(M^{-1})$ This is the celebrated type I see-saw mechanism.
In this paper we assume $M_3\gg 10^{14}\,{\rm GeV}$ so that $N_3$ 
decouples from the seesaw and effectively a two RH neutrino model ($n=2$) is recovered.

Let us now introduce the relevant general quantities for leptogenesis.
First of all notice that in addition to  $M_3\gg 10^{14}\,{\rm GeV}$,
we will also impose the condition $M_2\ll 10^{12}\,{\rm GeV}$.
In this case lepton flavour effects have to be
taken into account at the asymmetry production
from $N_1$ and $N_2$ decays \cite{flavoreffects}.

This can be done calculating the total $B-L$ asymmetry as the sum,
\be
N_{B-L}= \sum_{\alpha=e,\m,\t} \, N_{\D_{\a}}  \, ,
\ee
of the flavoured asymmetries defined as $\D_{\a}\equiv B/3 - L_{\a}$.
Notice that $N_X$ indicates any particle number or asymmetry $X$
calculated in a portion of co-moving volume containing one heavy neutrino
in ultra-relativistic thermal equilibrium, i.e.\ such that $N^{\rm eq}_{N_i}(T\gg M_i)=1$.
The baryon-to-photon number ratio at recombination is then given by
\be
\eta_B= a_{\rm sph}\,{N_{B-L}^{\rm f}\over N_{\gamma}^{\rm rec}}\simeq 0.01\,N_{B-L}^{\rm f} \, ,
\ee
to be compared with the  value measured from the CMB anisotropies observations \cite{WMAP7}
\be\label{etaBobs}
\eta_B^{\rm CMB} = (6.2 \pm 0.15)\times 10^{-10} \, .
\ee
The $N_i$ RH neutrinos' decay widths are given by
\be
\widetilde{\G}_i \equiv \Gamma_{i}+\bar{\G}_{i}= 
{M_i\,(m^{\dagger}\,m_D)_{ii}\over 8\,\pi\,v^2} \, ,
\ee
where $\Gamma_{i}$ and $\bar{\G}_{i}$ are respectively the $N_i$ decay rates into leptons
and anti-leptons (at zero temperature).
The key quantities encoding the main features of the kinetic evolution are
the decay parameters defined as \cite{kt,annals}
\be \label{K}
K_i=\frac{\widetilde{\G}_i}{H(T=M_i)}={\mti \over m_{\star}} \, , \hspace{10mm} {\rm where} \hspace{10mm} \mti
\equiv{(m_D^{\dagger}\,m_D)_{ii} \over M_i}
\ee
are the effective neutrino masses \cite{plumacher} and $m_{\star}$ is
the (SM) equilibrium neutrino mass defined by \cite{orloff,annals}
\begin{equation}\label{d}
m_{\star}\equiv {16\, \pi^{5/2}\,\sqrt{g^{*}_{SM}} \over 3\,\sqrt{5}}\, {v^2 \over M_{\rm Pl}} \simeq 1.08\times
10^{-3}\,{\rm eV} \, .
\end{equation}
The flavour composition of the lepton flavour quantum 
states produced by the $N_i$ decays can be written as,
\be
|{\ell}_i\rangle =
\sum_{\a}\,{\cal C}_{i\a}\,|{\ell}_\a \rangle \, , \;\;\;\;
{\cal C}_{i\a} \equiv  \langle {\ell}_\a|\ell_i \rangle  \;\;\;\; (i=1,2 \, , \a=e,\mu,\t) \,
\ee
and
\be
|\bar{\ell}_i\rangle =
\sum_{\a}\,\bar{{\cal C}}_{i\a}\,|\bar{{\ell}}_\a \rangle \, , \;\;
\bar{{\cal C}}_{i\a} \equiv  \langle \bar{{\ell}}_\a|\bar{\ell}_i \rangle \, , \;\; (i=1,2 \, , \a=e,\mu,\t) \, ,
\ee
where we notice that in general the
final anti-lepton states produced by the $N_2$ decays are not in general
the CP conjugated of the final lepton states and therefore, in general,
${\cal C}^*_{i\a} \neq \bar{{\cal C}}_{i\a}$ \cite{flavoreffects}.
Only at tree level one has ${\cal C}^*_{i\a} = \bar{{\cal C}}_{i\a}$ and in this case,
in terms of the Dirac mass matrix, they are given by 
\be\label{C0mD}
 {\cal C}^0_{i\alpha} =   {m^{\star}_{D\a i} \over \sqrt{(m_D^{\dagger}\,m_D)_{ii}}}
 \hspace{6mm} \mbox{\rm and} \hspace{6mm}
 \bar{\cal C}^0_{i\alpha} = { m_{D\a i}\over 
 \sqrt{(m_D^{\dagger}\,m_D)_{ii}}}  \, .
 \ee
Introducing the branching ratios
$P_{i\a} \equiv |{\cal C}_{i\a}|^2$ and  $\bar{P}_{i\alpha}\equiv |\overline{{\cal C}}_{i\a}|^2$ respectively,
i.e. the probabilities that a lepton or an anti-lepton is measured in
the $\alpha$ light lepton flavour eigenstate,
we can define the flavoured decay rates $\Gamma_{i\a}\equiv P_{i\a}\,\G_i$ and $\bar{\Gamma}_{i\a}\equiv \bar{P}_{i\a}\,\bar{\G}_i$.
In a three flavoured regime, when the produced lepton quantum states rapidly collapse into an incoherent mixture
of flavour eigenstates, the $\Gamma_{i\a}$ and the $\bar{\Gamma}_{i \a}$ can be identified
with the flavoured decay rates into $\a$ leptons, $\Gamma (N_i \ra \phi^\dagger \, l_\alpha)$ ,
and anti-leptons, $\Gamma (N_i \ra \phi \, \bar{l}_\alpha)$, respectively and
the $P_{i\a}$ and $\bar{P}_{i\a}$ with their branching ratios.
The total flavoured decay rates (at zero temperature) are then given by
\be
\widetilde{\G}_{i\a}\equiv \G_{i\a}+\bar{\G}_{i\a}= {M_i\,|m_{D\a i}|^2 \over 8\,\pi\,v^2} \,.
\ee
Correspondingly, the flavoured decay parameters are defined as
\be\label{Kial}
K_{i\a}\equiv {\tilde{\G}_{i\a}\over H(T=M_i)}= {|m_{D\a i}|^2 \over M_i \, m_{\star}} \simeq P^0_{i\a}\,K_i \, ,
\ee
where $P^0_{i\a}\equiv (P_{i\a}+\bar{P}_{i\a})/2 $ are the tree level branching ratios.

We can then define the flavored CP asymmetries as
\be
\ve_{i\a}\equiv
\frac{\Gamma_{i\a} - \overline{\Gamma}_{i\a}}{\Gamma_i + \bar{\Gamma}_i} \, ,
\ee
so that for the total $C\!P$ asymmetries one has
\be
\ve_i\equiv  {\G_i-\bar{\G}_i\over \G_i+\bar{\G}_i} =\sum_\a \ve_{i\a} \,.
\ee
%In the flavour basis, where the charged lepton Yukawa matrix $Y_e$ and the RH neutrino
%mass matrix $M$ are both diagonal with real and positive eigenvalues, the neutrino Yukawa
%matrix $Y_{\nu}$ will have elements denoted as $h_{\alpha i}$.
The flavored $C\!P$ asymmetries can then be calculated using \cite{crv}
\bea\label{epsia}
\ve_{i\a} & = & - \frac{3}{16 \p\,v^2\,(m_D^{\dag}\,m_D)_{ii}}
\sum_{j\neq i} \left\{ {\rm Im}\left[m_{D\a i}^{\star} m_{D\a j}(m_D^{\dag}\,m_D)_{i j}\right]
\frac{\x(x_j/x_i)}{\sqrt{x_j/x_i}}+ \right. \\  & &  \nonumber
\left. \frac{2}{3(x_j/x_i-1)}{\rm Im}
\left[m_{D\a i}^{\star}\,m_{D\a j}(m_D^{\dag}\,m_D)_{j i}\right]\right\} \, ,
\eea
where we defined $x_i \equiv \left({M_i/M_1}\right)^2$ and
\be\label{xi}
\xi(x)= {2\over 3}\,x\, \left[(1+x)\,\ln\left({1+x\over x}\right)-{2-x\over 1-x}\right] \, .
\ee
The second term on the right-hand side
in the expression for
$\ve_{i\a}$ due to the self-energy diagram has so far been neglected  in studies
of $N_1$ leptogenesis within the 2 RH neutrino model except in \cite{bounds} where however the 2 RH neutrino model
was not the main focus. In \cite{dirac} it was noticed that this term
could play a relevant role in the calculation of the heavier $N_2$ RH neutrino asymmetry,
and it will be included in our analysis
\footnote{It was discussed recently in \cite{Antusch:2009gn} that this term is generically dominant in scenarios where two RH neutrinos form a quasi-Dirac pair. Its size is related to the non-unitarity of the leptonic mixing matrix, caused by an effective dimension 6 operator. Since this scenario implies $M_1 \simeq M_2$, it will not be studied here.}.
%This will be the novel point of our analysis: an account of the second term in the $\ve_{i\a}$
%jointly with the asymmetry produced from the next-to-lightest RH neutrinos within the two RH neutrino model.

If we also introduce the variables
\be
z\equiv {M_1 \over T} \,  \hspace{5mm}
\mbox{\rm and} \hspace{5mm}
z_i \equiv \sqrt{x_i} \, z \, ,
\ee
the decay and the washout terms can be written respectively as
\be
D_i(z) = K_i\,x_i\,z\,
\left\langle {1\over\gamma_i} \right\rangle  \;\;\;\;\; \mbox{\rm and} \;\;\;\;\;
W_i(z) = {1\over 4}\,K_i \, \sqrt{x_i}\,{\cal K}_1(z_i)\,z_i^3 \, ,
\ee
where the averaged dilution factors can be expressed in terms of the Bessel functions,
$\left\langle {1/\gamma_i} \right\rangle = {{\cal K}_{1}(z_i) / {\cal K}_{2}(z_i)}$.

%---------------------------------------------------------------------------------
\section{An analytical solution  for the final asymmetry from Boltzmann equations}
%---------------------------------------------------------------------------------

After having set up the notation and framework, in this Section we write down the
Boltzmann equations and give an analytical solution for the final asymmetry.
We will  impose throughout the paper $M_2 \gtrsim 3\,M_1$, so that the hierarchical
(non-resonant) limit holds
and $\xi\simeq 1$ (cf. eq.~(\ref{xi})) is a good approximation \cite{beyond}.
In this limit the production and the wash-out  from the $N_2$'s and the production and the wash-out from the $N_1$'s can be treated as
two separate stages. In a first stage, the
 asymmetry is produced and washed-out by the $N_2$ processes, while in a second
 stage when the asymmetry is produced and washed-out by $N_1$ processes.

%At the same time we can also assume the hierarchical limit for the $C\!P$ asymmetries,
%neglecting the enhancement from the self energy contribution and using the approximation $\xi\simeq 1$.

\subsection{Production of the asymmetry from $N_2$ processes}
%%%%%%%%%%%%%%%%%%%%%%%%%%%%%%%%%%%%%%%%%%%%%%%%%%%%%%%%%%%%%
Recall that in the SM, if leptogenesis occurs at
temperatures $T\sim M_1$, where $M_1$ is the mass of the lightest RH neutrino,
then one has to distinguish two possible cases. If
 $10^5 \: \mbox{GeV} \ll M_1 \ll 10^{9} \: \mbox{GeV}$,
then charged $\mu$ and $\tau$ Yukawa interactions are in thermal equilibrium
and all flavours in the Boltzmann equations are to be
treated separately. For
$10^9 \: \mbox{GeV} \ll M_1 \ll 10^{12} \: \mbox{GeV}$, only the $\tau$ Yukawa interactions
are in equilibrium and are treated separately in the Boltzmann equations, while
the $e$ and $\mu$ flavours are indistinguishable.

In the case of $N_1$ leptogenesis, it is well known that
the dominant contribution from the first term on the right-hand side of eq.~(\ref{epsia})
to the flavoured  $C\!P$ asymmetries $\ve_{1\a}$ in the eq.~(\ref{epsia}) is bounded from above,
leading to a lower bound $M_{1}\gtrsim 10^9\,$GeV
in order for the $C\!P$ asymmetries to be sufficiently large.
We will see later that this conclusion also applies when the
production from the next-to-lightest RH neutrino $N_2$ is taken into account
and when both terms in the eq.~(\ref{epsia}), both for $\ve_{1\a}$ and $\ve_{2\a}$, are taken into account as well.

Assuming $M_{1}\gtrsim 10^9\,$GeV, we are always in the
two-lepton flavour regime, where only the tauon charged
lepton interactions are fast enough to break the coherent evolution of the final leptons.
This implies that we have to track separately the asymmetry in the tauon flavour and the
asymmetry in the flavour $\g_2$ defined as the coherent superposition of the
electron and muon components in the lepton quantum states
$|{\ell}_2\rangle$ produced by $N_2$ decays, explicitly
\be
|{\ell}_{\g_2}\rangle = 
{{\cal C}_{2e}\over \sqrt{|{\cal C}_{2e}|^2 + |{\cal C}_{2\mu}|^2}} \,|{\ell}_e \rangle +
{{\cal C}_{2\mu}\over \sqrt{|{\cal C}_{2e}|^2 + |{\cal C}_{2\mu}|^2}}\,|{\ell}_\mu \rangle  \, ,
\ee
and in the anti-lepton quantum states $|\bar{\ell}_{\g_2}\rangle$ analogously defined.

We neglect the coupling between the dynamics of distinct flavours $\a \neq \b$,
in the specific case of the two flavours $\t$ and $\g_2$ \cite{us}.
Therefore, in the stage where the asymmetry is produced by $N_2$ decays, the relevant kinetic
equations can be written as
 \bea
{dN_{N_2}\over dz} & = & -D_2\,(N_{N_2}-N_{N_2}^{\rm eq}) \, ,\\
{dN_{\D_{\g_2}}\over dz} & = &
-\ve_{2\g}\,D_2\,(N_{N_2}-N_{N_2}^{\rm eq})- P_{2\g}^{0}\,W_2\,N_{\D_{\g_2}} \, ,\\
{dN_{\D_{\t}}\over dz} & = &
-\ve_{2\t}\,D_2\,(N_{N_2}-N_{N_2}^{\rm eq})- 
P_{2\t}^{0}\,W_2\,N_{\D_{\tau}
}  \, ,
\eea
where we defined $\ve_{2\g}\equiv \ve_{2e}+\ve_{2\mu}$ and $P^0_{2\g}\equiv P^0_{2e}+P^0_{2\m}$.
This set of classical Boltzmann equations neglects different effects
that have been studied in the last years such as thermal masses \cite{thermal},
decoherence  \cite{flosc}, quantum kinetic effects \cite{qke}, momentum dependence \cite{momentum}, flavour coupling \cite{us}.
In the strong wash-out regime ($K_{2e}+K_{2\m},K_{2\t}\gtrsim 5$)
these effects give at most ${\cal O}(1)$ factor corrections.

The two asymmetries $N_{\D_{\a}}\,(\a=\t,\g_2)$  freeze out
at $T=M_2/z_{B\a}$ where \cite{beyond}
\be z_{B}(K_{i\a}) \simeq
2+4\,K_{i\a}^{0.13}\,e^{-{2.5\over K_{i\a}}}={\cal O}(1\div 10) \, .
\ee
At the end of the $N_2$ production stage, at $z \simeq z_{B2}\equiv {\rm max}[z_{B}(K_{2\g}),z_{B}(K_{2\t})]$,
one has
\be N_{\D_{\g_2}}^{z_{B2}}  \simeq  -\ve_{2\gamma}\,\k_{2\g} , \;\;
\hspace{5mm}\mbox{\rm and}\hspace{5mm}
N_{\D_{\tau}}^{z_{B2}}  \simeq  -\ve_{2\tau}\,\k_{2\tau} \, .
\ee
In the case of an initial thermal abundance, the efficiency
factors at the production are approximately given by \cite{annals,beyond,flavorlep}
\be \label{thermal}
\k_{i\a}\simeq \k(K_{i\a}) \equiv \frac{2}{K_{i\a} \, z_{\rm B}(K_{i\a})}\left[1-{\rm exp}\left(-\frac{1}{2} K_{i\a}\, z_{\rm
B}(K_{i\a})\right)\right]\, .
\ee
In the case of vanishing initial abundances, the efficiency
factors are the sum of two different contributions, a negative and a positive one, explicitly
\be \label{vanishing}
\k_{i\a} \simeq \k_{-}(K_i,P_{i\a}^{0})+ \k_{+}(K_i,P_{i\a}^{0}) \, .
\ee
The negative contribution arises from a first stage where 
$N_{N_i}\leq N_{N_i}^{\rm eq}$, for $z_i\leq z_i^{\rm eq}$,
and is given approximately by
\be\label{k-}
\k_{-}(K_i,P_{i\a}^{0})\equiv  -{2\over P_{i\a}^{0}}\
e^{-{3\,\pi\,K_{i\a} \over 8}} \left(e^{{P_{i\a}^{0}\over 2}\,N_{N_i}(z_{\rm eq})} - 1 \right) \, .
\ee
The positive contribution arises from a second stage where $N_{N_i}\geq N_{N_i}^{\rm eq}$, for $z_i\geq z_i^{\rm eq}$,
and is approximately given by
\be\label{k+} \k_{+}(K_i,P_{i\a}^{0})\equiv {2\over z_B(K_{i\a})\,K_{i\a}}
\left(1-e^{-{K_{i\a}\,z_B(K_{i\a})\,N_{N_i}(z_{\rm eq})\over 2}}\right) \, .
\ee
The $N_i$ abundance at $z_i^{\rm eq}$ is well approximated by the expression
\begin{equation}\label{nka}
N_{N_i}(z_i^{\rm eq}) \simeq  {N(K_i)\over\left(1 + \sqrt{N(K_i)}\right)^2}\, ,
\end{equation}
that interpolates between the limit $K_i\gg 1$, where $z_i^{\rm eq}\ll 1$ and $N_{\rm N_i}(z_i^{\rm eq})=1$, and the
limit $K_i\ll 1$, where $z_i^{\rm eq}\gg 1$ and $N_{N_i}(z_i^{\rm eq})=N(K_i)\equiv 3\p K_i/4$.
We will present all results for vanishing initial abundances, since this is the 
case with lower efficiency and that therefore 
yields the most stringent constraints. Notice, however, that for most of the
allowed regions, the strong wash-out
regime ($K_{i\a}\gg 1$) holds. In this case 
the efficiency factors coincide asymptotically  in the two cases since
they become independent of the initial  RH  neutrino abundances.

\subsection{Production and wash-out of the asymmetry from $N_1$ processes}
%--------------------------------------------------------------------------

When inverse $N_1$ processes start to be active at $z\simeq 1$, they break the coherent evolution of the
$|{\ell}_{\gamma_2}\rangle$ quantum states \cite{N2importance}. We describe this decoherent effect in terms of a
simple collapse of the quantum state neglecting decoherence effects that would be described by a density matrix approach.
This is justified since, thanks to the condition of hierarchical masses $M_2 \gtrsim 3\,M_1$,
the $N_2$-decays are already switched off at this stage and they do not interfere
with  $N_1$ inverse processes. The stage of decoherence
can then be regarded as a transient stage with no relevant consequences on the final asymmetry.

Therefore, at $T\sim M_1$, the $|{\ell}_{\g_2}\rangle$ quantum states can be described as an incoherent mixture
of a ${\ell}_{\g_1}$ component and of a  ${\ell}_{\g_1^{\bot}}$ component.
Both components are a coherent superposition of electron and muon flavour eigenstates.
The first has a flavour composition given by the projection of $|{\ell}_1\rangle$
on the $e-\m$ plane, while the second  is the projection
of the $|{\ell}_1\rangle$ orthogonal component on the $e-\m$ plane.
Analogously to the $|{\ell}_{\g_2}\rangle$, we can explicitly write 
the  $|{\ell}_{\g_1}\rangle$ as
\be
|{\ell}_{\g_1}\rangle = 
{{\cal C}_{1e}\over \sqrt{|{\cal C}_{1e}|^2 + |{\cal C}_{1\mu}|^2}} \,|{\ell}_e \rangle +
{{\cal C}_{1\mu}\over \sqrt{|{\cal C}_{1e}|^2 + |{\cal C}_{1\mu}|^2}}\,|{\ell}_\mu \rangle  \, .
\ee

Let us now define the probability
$p_{12}\equiv |\langle {\ell}_{\gamma_1} |{\ell}_{\gamma_2}\rangle|^2$. Using the
eqs.~(\ref{C0mD}), this can be
calculated from the Dirac mass matrix as
\be \label{p12}
p_{12}= {\left| {\cal C}^{\star}_{1e}\,{\cal C}_{2e} + 
{\cal C}^{\star}_{1\mu}\,{\cal C}_{2\mu} \right|^2 
\over \left(|{\cal C}_{1e}|^2 + |{\cal C}_{1\mu}|^2 \right)\,
         \left(|{\cal C}_{2e}|^2 + |{\cal C}_{2\mu}|^2 \right)} = 
 {1 \over P_{1\g}^{0}\,P_{2\g}^{0} } \, {|\sum_{\alpha=e,\m} (m^{\star}_{D\a 1}\,m_{D\a 2})|^2\over
(m^{\dagger}_D\,m_D)_{11}\,(m^{\dagger}_D\,m_D)_{22}}  \, .
\ee
Within the adopted kinetic description,  only the component
$\langle {\ell}_{\gamma_1} |{\ell}_{\gamma_2}\rangle\, |{\ell}_{\gamma_1}\rangle$ 
of $|{\ell}_{\gamma_2}\rangle$ interacts with the Higgs
in an inverse process producing $N_1$.
The  ${\ell}_{\gamma_1}$-orthogonal component 
$\langle {\ell}_{\gamma_1^{\bot}} |{\ell}_{\gamma_2}\rangle \, |{\ell}_{\gamma_1^{\bot}}\rangle=|{\ell}_{\gamma_2}\rangle - \langle {\ell}_{\gamma_1} |{\ell}_{\gamma_2}\rangle\, |{\ell}_{\gamma_1}\rangle$, is untouched. 
Analogous considerations hold for the anti-lepton quantum states.
In this way only the asymmetry in the flavour $\g_1$, that we indicate with $N_{\D_{\g_1}}$ is washed out,
while the asymmetry $N_{\D_{\g_1^\bot}}$ is not changed by $N_1$ inverse processes.

Therefore, under the action of $N_1$ decays and inverse processes, the $|{\ell}_{\gamma_2}\rangle$
quantum states collapse into an incoherent mixture of $|{\ell}_{\gamma_1}\rangle$
and  $|{\ell}_{\gamma_1^{\bot}}\rangle$ quantum states and analogously the $|\bar{\ell}_{\gamma_2}\rangle$.
Correspondingly, one has to calculate separately the two contributions $N_{\D_{\g_1}}$
and  $N_{\D_{\g_1^{\bot}}}$ to the asymmetry, in addition to the  $N_{\D_{\t}}$ asymmetry
in the tauon flavour.

Therefore, in this stage the set of Boltzmann equations is given by
\bea\label{flke}
{dN_{N_1}\over dz} & = & -D_1\,(N_{N_1}-N_{N_1}^{\rm eq}) \, ,\\
{dN_{\D_{\g_1}}\over dz} & = & -\ve_{1\g}\,D_1\,(N_{N_1}-N_{N_1}^{\rm eq})-
P_{1\g}^{0}\,W_1\,N_{\D_{1\g}} \, ,\\
{dN_{\D_{\g_1^\bot}}\over dz} & = & 0 ,\\
{dN_{\D_{\t}}\over dz} & = &
-\ve_{1\t}\,D_1\,(N_{N_1}-N_{N_1}^{\rm eq})- P_{1\t}^{0}\,W_1\,N_{\D_{\tau}}  \, ,
\eea
implying that $N_{\D_{\g_1^\bot}}$ remains constant.
Notice that effectively we have in the end, because of the heavy
flavour interplay, a three flavour regime where the final $B-L$ asymmetry can be calculated as
the sum of three contributions,
\be
N_{B-L}^{\rm f}= N_{\D_{\t}}^{\rm f}+N_{\D_{\g_1}}^{\rm f}+N_{\D_{\g_1^\bot}}^{\rm f} \, ,
\ee
where
\bea
N_{\D_{\g_1}}^{\rm f} & \simeq & - p_{12}\,\ve_{2\gamma}\,\k_{2\g}\,e^{-{3\,\pi\,\over 8}K_{1\gamma} }
-\ve_{1\gamma}\,\k_{1\g} \, , \\
N_{\D_{\g_1^\bot}}^{\rm f} & \simeq & -\,(1-p_{12})\,\ve_{2\gamma}\,\k_{2\g} \, , \\
N_{\D_{\tau}}^{\rm f} & \simeq &
-\ve_{2\tau}\,\k_{2\tau} \,e^{-{3\,\pi\,\over 8}K_{1\tau} } - \ve_{1\tau}\,\k_{1\tau} .
\eea
It is useful for our discussion to split the final asymmetry
into a contribution from $N_1$ decays and into a contribution from
$N_2$ decays,
\be\label{NBmLf}
N_{B-L}^{\rm f}= N_{B-L}^{\rm f (1)} + N_{B-L}^{\rm f (2)} \, ,
\ee
where
\be\label{NBmLf1}
N_{B-L}^{\rm f (1)} \simeq -\ve_{1\gamma}\,\k_{1\g} - \ve_{1\tau}\,\k_{1\tau}
\ee
and
\be\label{NBmLf2}
N_{B-L}^{\rm f (2)} \simeq - p_{12}\,\ve_{2\gamma}\,\k_{2\g}\,e^{-{3\,\pi\,\over 8}K_{1\gamma} } -
(1-p_{12})\,\ve_{2\gamma}\,\k_{2\g} - \ve_{2\tau}\,\k_{2\tau}\,
e^{-{3\,\pi\,\over 8}\, K_{1\tau} } \, .
\ee
In this way we clearly distinguish
the effect of taking into account the asymmetry produced from the next-to-lightest RH neutrinos $N_2$,
which has been neglected in 
previous analyses where the impact of the flavour structure on leptogenesis was studied in 2 RH neutrino models.

%%%%%%%%%%%%%%%%%%%%%%%%%%%%%%%%%%%%%%%%%%%%%%%%%%%%%%%%%%%%%%%%%%%%%%%%%%%%%%%%%%%%
\section{Combining the low energy neutrino data with the orthogonal parametrization}
%%%%%%%%%%%%%%%%%%%%%%%%%%%%%%%%%%%%%%%%%%%%%%%%%%%%%%%%%%%%%%%%%%%%%%%%%%%%%%%%%%%%

In this section we recast our expression for the final asymmetry in the orthogonal
parametrization, which provides a convenient way to connect the constraints from leptogenesis
to the information from current low energy neutrino experiments and the additional parameters
from the RH neutrino sector.

\subsection{Orthogonal parametrization for the two RH neutrino model}\label{sec:orthparam}
%%%%%%%%%%%%%%%%%%%%%%%%%%%%%%%%%%%%%%%%%%%%%%%%%%%%%%%%%%%%%%%%%%%%%

The light and heavy neutrino mass matrices can be diagonalized
by unitary matrices $U$ and $U_M$, respectively. Hence we have the relations
$U^{\dagger}m_{\nu}U^*=D_k$ and $U_M^{\dagger}MU_M^*=D_M$, where $D_k={\rm diag}(m_1, m_2, m_3)$
and $D_M={\rm diag}(M_1, M_2, M_3)$ are diagonal matrices containing the light and heavy
neutrino mass eigenvalues for three RH neutrinos.
In the basis where $Y_e$ is diagonal we identify
$U$ as the PMNS matrix. From above, one obtains,
\be
U^{\dagger}m_DM^{-1}m_D^TU^*=D_k \,.
\ee
Substituting $U_M^{\dagger}MU_M^*=D_M \label{eq:Md1}$ in the above equation we get,
\be
U^{\dagger}\, m_D\, U_M^*\, D_M^{-1}\, U_M^{\dagger}\, m_D^T\, U^*=D_k \,.
\ee
The $R$ matrix is defined as \cite{CasasIbarra}
\footnote{In terms of PMNS mixing matrix
$R=D_{\sqrt{M}}^{-1}U_M^{\dagger}m_D^TU^*_lU_{PMNS}^*D_{\sqrt{k}}^{-1}$
where $U_{PMNS}=U_l^{\dagger}U$.}
\be
R=D_{\sqrt{M}}^{-1}U_M^{\dagger}m_D^TU^*D_{\sqrt{k}}^{-1} \, , \label{eq:Rmat}
\ee
where $R$ is a complex orthogonal matrix $R^TR={I}$.
Eq.~(\ref{eq:Rmat}) parametrizes the freedom in the Dirac matrix
$m_D$, for fixed values of $U$, $D_k$ and $D_M$, in terms of a
complex orthogonal matrix $R$.

From eq.~(\ref{eq:Rmat}), in the basis where $M$
and $Y_e$ are diagonal, $m_D$ parameterizes as:
\be\label{orthopara}
m_D\,D_{\sqrt{M}}^{-1}=U\, D_{\sqrt{k}}\, R^T \, , \label{R}
\ee
where
$D_{\sqrt{k}}={\rm diag}(m_1^{1/2},m_2^{1/2}, m_3^{1/2})$
and
$D_{\sqrt{M}}^{-1}={\rm diag}(M_1^{-1/2}, M_2^{-1/2}, M_3^{-1/2})$
for three RH neutrinos. To be completely explicit
we can write the Dirac matrix $m_D$ as $m_{D\a i}$ where
$\a=e, \mu , \tau$ labels the rows and $i=1,2,3$ labels
the columns corresponding to the three RH neutrinos
and then expand eq.~(\ref{R}) as:
\bea
\pmatrix{ m_{De1}M_1^{-1/2} & m_{De2}M_2^{-1/2}  & m_{De3}M_3^{-1/2} \cr
         m_{D\mu 1}M_1^{-1/2} & m_{D\mu 2}M_2^{-1/2}  & m_{D\mu 3}M_3^{-1/2} \cr
           m_{D\tau 1}M_1^{-1/2} & m_{D\tau 2}M_2^{-1/2}  & m_{D\tau 3}M_3^{-1/2}}
= \pmatrix{ U_{e1}m_1^{1/2} & U_{e2}m_2^{1/2}  & U_{e3}m_3^{1/2} \cr
         U_{\mu 1}m_1^{1/2} & U_{\mu 2}m_2^{1/2}  & U_{\mu 3}m_3^{1/2} \cr
           U_{\tau 1}m_1^{1/2} & U_{\tau 2}m_2^{1/2}  & U_{\tau 3}m_3^{1/2}} R^T.
\label{R_explicit1}
\eea
The eq.~(\ref{R_explicit1}) enables the Dirac matrix to be determined in terms of the completely free
parameters of the complex orthogonal matrix $R$, for a fixed physical parameter set $U,m_i,M_i$.
For example we can scan over the parameters of $R$ for a fixed $U,m_i,M_i$.

As remarked, the two RH neutrino model can be regarded as a limiting case of three RH neutrinos
where one of the RH neutrinos decouples from the see-saw mechanism either because it is very heavy or because
its Yukawa couplings are very weak \cite{King:1999mb}. In our case we shall consider the former situation $M_3
\rightarrow \infty$. Then we are left with only two non-zero physical neutrino masses which can be identified as either
$m_2, m_3$ with $m_1 \rightarrow 0$ for a normal hierarchy (NH), or $m_1, m_2$ with $m_3 \rightarrow 0$ for an inverted
hierarchy (IH).

For the case of two RH neutrinos of mass $M_1,M_2$, and
two physical neutrino masses $m_2, m_3$, for the case
of a normal hierarchy, with $m_1 \rightarrow 0$,
\bea \pmatrix{ m_{De1}M_1^{-1/2} & m_{De2}M_2^{-1/2}  \cr
         m_{D\mu 1}M_1^{-1/2} & m_{D\mu 2}M_2^{-1/2}   \cr
           m_{D\tau 1}M_1^{-1/2} & m_{D\tau 2}M_2^{-1/2} }
= \pmatrix{ U_{e2}m_2^{1/2}  & U_{e3}m_3^{1/2} \cr
          U_{\mu 2}m_2^{1/2}  & U_{\mu 3}m_3^{1/2} \cr
           U_{\tau 2}m_2^{1/2}  & U_{\tau 3}m_3^{1/2}} R^T.
\eea
For the case of two RH neutrinos of mass $M_1,M_2$,
and two physical neutrino masses $m_1, m_2$, for the case
of an inverted hierarchy, with $m_3 \rightarrow 0$,
\bea \pmatrix{ m_{De1}M_1^{-1/2} & m_{De2}M_2^{-1/2}  \cr
         m_{D\mu 1}M_1^{-1/2} & m_{D\mu 2}M_2^{-1/2}   \cr
           m_{D\tau 1}M_1^{-1/2} & m_{D\tau 2}M_2^{-1/2} }
= \pmatrix{ U_{e1}m_1^{1/2} & U_{e2}m_2^{1/2}  \cr
         U_{\mu 1}m_1^{1/2} & U_{\mu 2}m_2^{1/2}   \cr
           U_{\tau 1}m_1^{1/2} & U_{\tau 2}m_2^{1/2} } R^T.
\label{R_explicit2}
\eea
%%%%%%%%%%%%%%%%%%%%%%%%%%%%%%%%%%%%%%%%%%%%%%%%%%%%%%%%%%%%%%%%%%%%%%%%%%%%%%%%%%%%%%%%%%%%%%%%%%%%%%%%%%%%%%%%%%%%%%%%%%%%%%%%%%%%%%%%%%%%
In each case the $3\times 2$ Dirac mass matrix $m_{Dli}$ is parametrized
in terms of a $2\times 2$ complex R-matrix which can be written as:
\bea\label{R2RH}
R= \pmatrix{ \cos z & \zeta\,\sin z \cr
 -\sin z  & \zeta\,\cos z }
\eea
where $z$ is a complex angle and $\zeta=\pm 1$ accounts for the possibility of two different
choices (`branches').

On the other hand, if we consider the two RH neutrino model as a limit of the 3 RH neutrino model for $M_3\gg
10^{14}\,{\rm GeV}$, then the orthogonal $R$ matrix tends to
\bea\label{RNH}
R^{(NH)}= \pmatrix{ 0 & \cos z & \zeta\,\sin z \cr 0 & -\sin z  & \zeta\,\cos z \cr 1 & 0 & 0}
\eea
and
\bea\label{RIH}
R^{(IH)}= \pmatrix{
\cos z & \zeta\,\sin z & 0 \cr
 -\sin z  & \zeta\,\cos z & 0 \cr
0 & 0 & 1 } \, .
\eea
Notice that the two branches cannot be obtained from each other with a continuous variation
of the complex angle. This can be clearly seen if one considers the following general parametrization
of the orthogonal matrix as a product of three complex rotations,
\be\label{second}
R(z_{23},z_{13},z_{12})
%={\rm diag}(\zeta ',\zeta ',\zeta ')\,
=\zeta' R_{23}(z_{23})\,\,
 R_{13}(z_{13})\,\,
 R_{12}(z_{12})\,\, ,
\ee
where
\be\label{Rpara}
\mbox{\tiny $
R_{23}=
\left(
\begin{array}{ccc}
  1  &  0   & 0   \\
  0  & \cos z_{23}  & \sin z_{23} \\
  0 & -\sin z_{23}  & \cos z_{23}
\end{array}
\right) \,\,\, , \,
R_{13}=
\left(
\begin{array}{ccc}
\cos z_{13}  & 0 & \sin z_{13} \\
    0 & 1 & 0 \\
-\sin z_{13}  & 0 & \cos z_{13}
\end{array}
\right) \,\,\, , \,
R_{12}=
\left(
\begin{array}{ccc}
 \cos z_{12}  & \sin z_{12}          & 0 \\
-\sin z_{12}  & \cos z_{12} & 0 \\
  0 & 0 & 1
\end{array}
\right)
$
}
\ee
and where the overall sign $\zeta'=\pm 1$ takes into account the
possibility of a  parity transformation as well. Within this general
case the two RH neutrino model $R$ matrix for NH eq.~(\ref{RNH})
is obtained for $z_{13}= z$, $z_{23}=\zeta \,z_{12}= \pi/2$ and $\zeta'=\zeta$,
clearly showing that the two branches for $\zeta=\pm 1$
cannot be obtained from each other with a continuous variation of
the complex angle $z$ (analogously for IH). More simply, it is sufficient
to recognise that $\det R = \zeta$ and to notice that matrices $R$ with determinant
$\det R = 1$ cannot be continuously deformed into matrices $R$ with 
$\det R = -1 $.

We will refer in the following to this kind of view of the two RH neutrino model.
We can perform scans over $z$ for a fixed $U,m_i,M_i$.

Neutrino oscillation experiments measure two neutrino mass-squared differences.
In the case of the two RH neutrino model for NH one has $m_1=0$, $m_2=m_{\rm sol}\equiv \sqrt{\Delta m^2_{\rm sol}}=(0.00875\pm 0.00012)\,{\rm eV}$
and $m^{\,2}_3-m_2^{\,2}=\Delta m^2_{\rm atm}$. The heaviest neutrino
has therefore a mass $m_3=m_{\rm atm} \equiv \sqrt{\Delta m^2_{\rm atm}+\Delta m^2_{\rm sol}}=
(0.050\pm 0.001)\,{\rm eV}$ \cite{oscillations}.
In the case of IH one has $m_3=0$, $m_2=m_{\rm atm}$ and $m_1=\sqrt{m_{\rm atm}^2-m_{\rm sol}^2}$.

We will adopt the following parametrisation  for the matrix $U$
in terms of the mixing angles, the Dirac phase $\delta$ and
 the Majorana phase $\alpha_{21}$  \cite{PDG}
\begin{equation}\label{Umatrix}
U=\left( \begin{array}{ccc}
c_{12}\,c_{13} & s_{12}\,c_{13} & s_{13}\,e^{-{\rm i}\,\d} \\
-s_{12}\,c_{23}-c_{12}\,s_{23}\,s_{13}\,e^{{\rm i}\,\d} &
c_{12}\,c_{23}-s_{12}\,s_{23}\,s_{13}\,e^{{\rm i}\,\d} & s_{23}\,c_{13} \\
s_{12}\,s_{23}-c_{12}\,c_{23}\,s_{13}\,e^{{\rm i}\,\d}
& -c_{12}\,s_{23}-s_{12}\,c_{23}\,s_{13}\,e^{{\rm i}\,\d}  &
c_{23}\,c_{13}
\end{array}\right)
\cdot {\rm diag}\left( 1, e^{i\,{\a_{21}\over 2}}, 1\right)\,
\end{equation}
and the following $2\,\sigma$ ranges for the three mixing angles \cite{oscillations}
\be\label{twosigma}
\theta_{12}= (31.3^\circ-36.3^\circ)  \, , \;\;\;
\theta_{23}= (38.5^\circ-52.5^\circ) \, , \;\;\;
\theta_{13}= (0^\circ-11.5^\circ) \, .
\ee

As we will see, there will be some sensitivity to the low energy neutrino parameters,
in particular to the value of $\theta_{13}$, of the Dirac phase and of the Majorana phase.
We will therefore perform the scans with the following 4 benchmark $U_{PMNS}$ choices A,B,C and D:
\be\label{bench1}
A: \hspace{.5cm} \theta_{13} =  0            , \,     \delta=0 , \, \frac{\a_{21}}{2}=0
\ee

\be\label{bench2}
B: \hspace{.5cm} \theta_{13} = 11.5^{\circ}  , \,     \delta=0  , \, \frac{\a_{21}}{2}=0
\ee

\be\label{bench3}
C: \hspace{.5cm}\theta_{13} = 11.5^{\circ}  , \,      \delta={\pi\over 2}   , \, \frac{\a_{21}}{2}=0
\ee

\be\label{bench4}
D: \hspace{.5cm} \theta_{13} = 11.5^{\circ}  , \,     \delta=0  , \, \frac{\a_{21}}{2}= {\pi\over 2},
\ee
where for all benchmarks the solar mixing angle and the atmospheric mixing angle are fixed to
$\theta_{12}=34^{\circ}$ and $\theta_{23}=45^{\circ}$ which are chosen to be close to their
best fit values.
Notice that benchmark A is close to tri-bimaximal (TB) mixing \cite{tribimaximal}, with no low energy CP violation in the
Dirac or Majorana sectors,
while the remaining benchmarks all feature the highest allowed reactor angle consistent with the
recent T2K electron appearance results \cite{:2011sj}.
On the other hand, varying the atmospheric and solar angles within their experimentally allowed ranges
has little effect on the results, so all benchmarks have the fixed atmospheric and solar angles above.
Benchmark B involves no CP violation in the low energy Dirac or Majorana sectors, with any CP violation
arising from the high energy see-saw mechanism parametrized by the complex angle $z$.
Benchmark C involves maximal low energy CP violation via the Dirac  phase, corresponding to the
oscillation phase $\delta = \pi/2$, but has zero low energy CP violation via the Majorana phase, with $\alpha_{21}/2=0$.
Benchmark D involves maximal CP violation from the Majorana sector, $\alpha_{21}/2=\pi/2$, but zero CP violation
in the Dirac sector, $\delta =0$. These benchmark points are thus chosen to span the relevant
parameter space and to illustrate the effect of the different sources of CP violation.

\subsection{Decay parameters in the orthogonal parametrization}
%%%%%%%%%%%%%%%%%%%%%%%%%%%%%%%%%%%%%%%%%%%%%%%%%%%%%%%%%%%%%%

We can start first expressing the quantities $\mti$, $K_i$, $K_{i\alpha}$ and $p_{12}$ in the orthogonal
parametrization. For the effective neutrino masses and the total decay parameters one has simply
\be\label{mtiR}
\mti=\sum_k\,m_k\,|R_{ik}|^2 \, , \hspace{5mm}\mbox{\rm and} \hspace{5mm} K_i= \sum_k\,{m_k\over m_{\star}}\,|R_{ik}|^2 \, .
\ee
Substituting
$m_{D_{\a i}}=\sqrt{M_i} \,\sum_{k}\,\sqrt{m_k} \,U_{\a k}\,R_{ik}$ 
(cf. eq.~(\ref{orthopara})) into
$K_{i\a}=\left|\,m_{D_{\a i}}\,\right|^2 / (M_i\,m^{*})$, one obtains
\be
K_{i\a}={1 \over m^{*}} \, \left| \, \sum_{k}\,\sqrt{m_k} \, U_{\a k} \, R_{ik} \, \right|^2 \, .
\ee
From this expression and from the definition of $K_{i\a}$ in eq.~(\ref{Kial}),
one then also obtains
\be\label{P01a}
P_{i\a}^0 = \frac{\left|\sum_{k'} \sqrt{m_{k'}} \,U_{\a k'}\,R_{ik'} \right|^2 }
{\widetilde m_{i}} .
\ee
Finally, substituting $m_{D_{\a i}}$ 
from eq.~(\ref{R_explicit1}) into eq.~(\ref{p12}) for $p_{12}$ yields
\be
p_{12}=
{1 \over P_{1\g}^{0} \,P_{2\g}^{0} \, \widetilde{m}_1\,  \widetilde{m}_2}\,\left|\,\sum_{k,k'}\sum_{\a=e,\m}\sqrt{m_k\,m_{k'}}\,U^{*}_{\a k}\,U_{\a k'}\,R^{*}_{1k}\,R_{2k'} \, \right|^2 \, .
\ee
With two RH neutrinos, we may express all quantities explicitly
in terms of complex angle $z$ for NH ($m_1=0$) as
\be
\mt = m_{\rm sol}\,|\cos z|^2 + m_{\rm atm}\,|\sin z|^2
\, , \hspace{5mm}  K_1 = K_{\rm sol}\,|\cos z|^2 + K_{\rm atm}\,|\sin z|^2 \, ,
\ee
\be
\mtt = m_{\rm sol}\,|\sin z|^2 + m_{\rm atm}\,|\cos z|^2
\, , \hspace{5mm}  K_2 = K_{\rm sol}\,|\sin z|^2 + K_{\rm atm}\,|\cos z|^2 \, ,
\ee
\be \label{K1alpha}
K_{1\a} ={1 \over m^{*}} \, \left| \sqrt{m_{\rm sol}}\,U_{\a 2}\,\cos z + \zeta\, \sqrt{m_{\rm atm}}\,U_{\a 3}\,\sin z \right|^2  \, ,
\ee
\be\label{K2alpha}
K_{2\a} ={1 \over m^{*}} \, \left| \zeta\,\sqrt{m_{\rm atm}}\,U_{\a 3}\,\cos z - \sqrt{m_{\rm sol}}\,U_{\a 2}\,\sin z  \right|^2  \, ,
\ee
%and
%\be \label{p12NH}
%p_{12} ={1 \over \pasquale{m^{*2}\,K_{1\gamma}\,K_{2\gamma}}}\left|m_{\rm atm} \, \cos %z \, \sin^{*}z - m_{\rm sol} \, \cos^{*}z \sin z \,\right|^2 \, ,
%\ee
where we defined $K_{\rm sol}\equiv m_{\rm sol}/m_{\star}\sim 10$ and $K_{\rm atm}\equiv m_{\rm atm}/m_{\star}\sim 50$.

For IH, ($m_3=0$), we can approximate $m_1 \simeq m_2 = m_{\rm atm}$ and simplify further
\be
\mt \simeq \mtt \simeq m_{\rm atm}\,\left(|\cos z|^2 + |\sin z|^2 \right)=m_{\rm atm}\, \cosh [2 {\rm Im}z]
\, ,
\ee
\be
K_1 \simeq K_2 \simeq K_{\rm atm}\,\left(|\cos z|^2 + |\sin z|^2 \right)=K_{\rm atm}\, \cosh [2 {\rm Im}z] \, ,
\ee
\be
K_{1\a} \simeq  K_{\rm atm} \, \left| U_{\a 1}\,\cos z + \zeta\,U_{\a 2}\,\sin z  \right|^2 \, ,
\ee
\be
K_{2\a} \simeq   K_{\rm atm} \, \left| \zeta\, U_{\a 2}\,\cos z - U_{\a 1}\,\sin z  \right|^2 \, .
\ee
%and
%\be\label{p12IH}
%p_{12}\simeq {4 \left|{\rm Im} 
%\left[\, \cos^{*}z \sin z \,\right]\right|^2 \over \pasquale{P^0_{1\g}\,P^0_{2\g}} \left( |%\cos z|^2 + |\sin z|^2 \right) ^2} \,
%= {\left| \tanh [2 {\rm Im}z] \right|^2\over \pasquale{P^0_{1\g}\,P^0_{2\g}}} \, .
%\ee
%The inequality $0 \leq \left| \tanh [2 {\rm Im}z] \right|^2 \leq 1$ is also satisfied for all z, as it must be given that $p_{12}(z)$ is a probability.

\begin{figure}
 \includegraphics [width=0.5\textwidth]{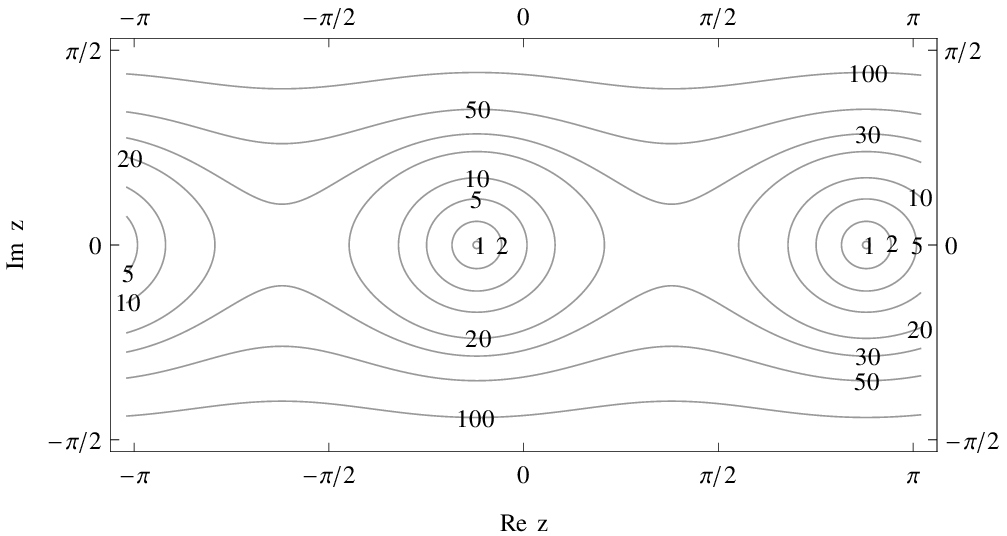}
 \includegraphics [width=0.5\textwidth]{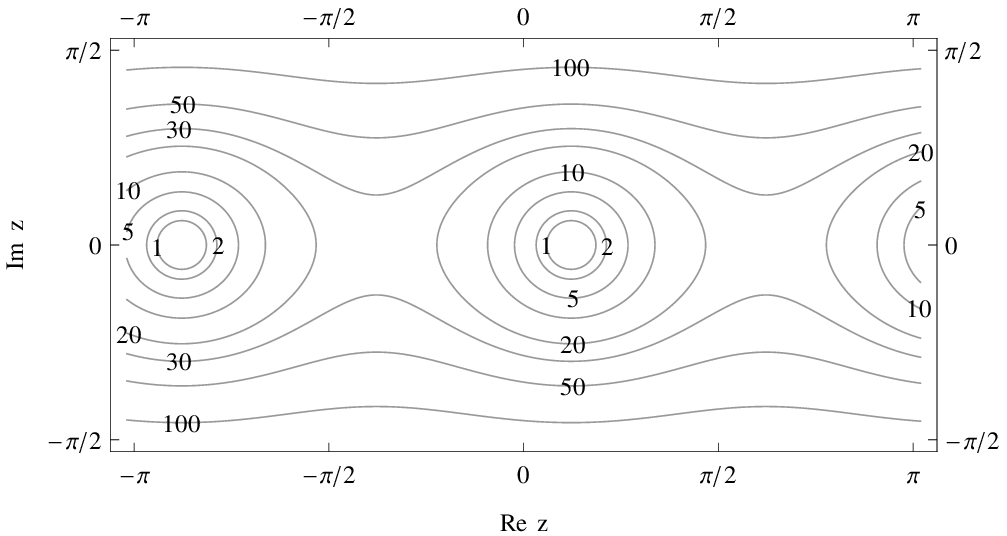}
 \includegraphics [width=0.5\textwidth]{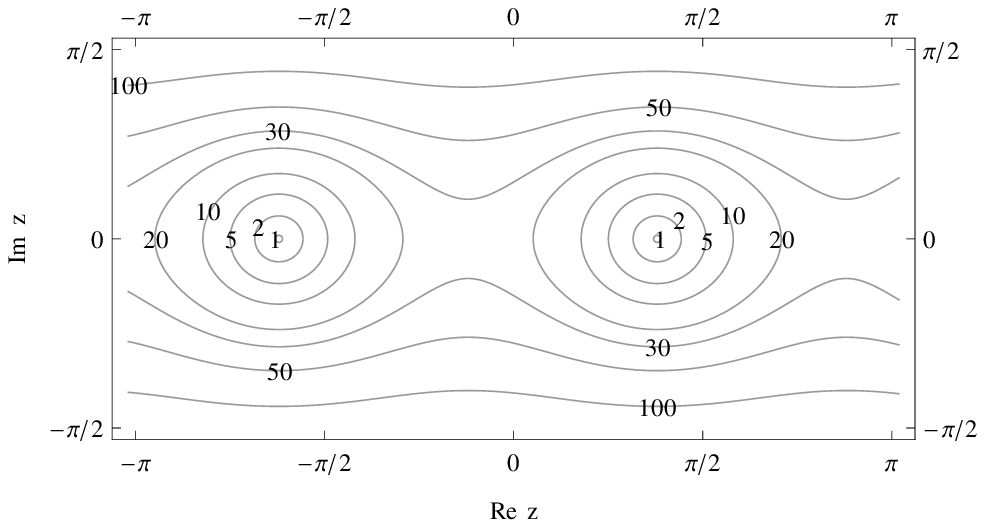}
 \includegraphics [width=0.5\textwidth]{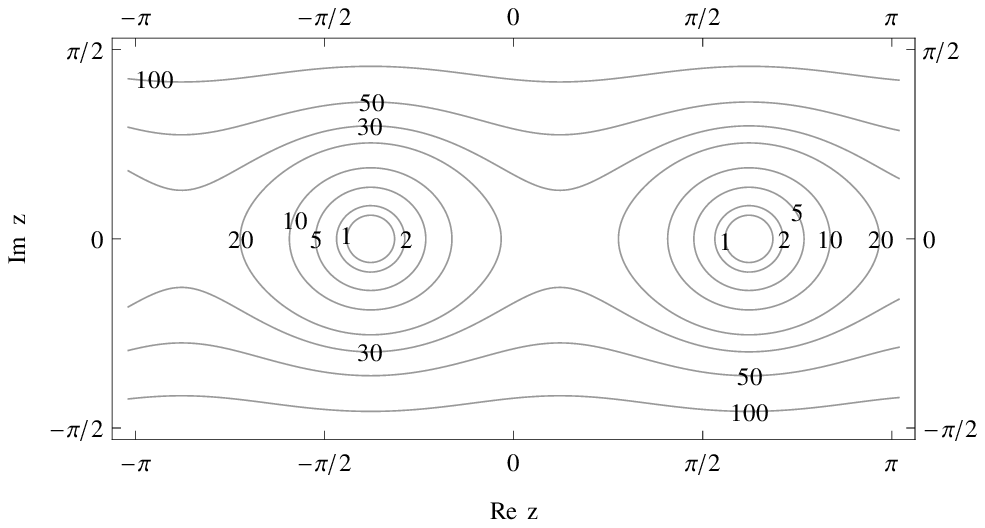}
 \vspace*{-7mm}
\caption{
Contour plots showing $K_{1\g} = \sum_{\a = e,\m} K_{1\a}$  (upper left panel), $K_{1\t}$ (upper right panel), 
$K_{2\g}= \sum_{\a = e,\m} K_{2\a}$ (lower left panel) 
and $K_{2\t}$ (lower right panel) dependence on complex angle z for benchmark B (cf. eq.~(\ref{bench2})), $\zeta=+1$, and NH.}
\label{fig:K1alpha}
\end{figure}

In Fig.~\ref{fig:K1alpha} we show contour plots of the flavoured decay parameters
$K_{1\g}, K_{1\t}$ and $K_{2\g}, K_{2\t}$ in the relevant region of the $z$ complex plane for NH and for the
benchmark case B, since this will prove the case maximizing the effect of the $N_2$ asymmetry production. Notice that Fig.~\ref{fig:K1alpha} is periodic in $\p$ along the $ {\rm Re}z$ axis as can be also inferred analytically from eqs.~(\ref{K1alpha}) and (\ref{K2alpha}) using double angle identities.
The most significant feature to be noticed at this stage is that for most of the parameter space $K_{i\a}\gg 1$ holds.
In these regions a strong wash-out regime is realized and this implies that the
 dependence of the results on the initial conditions is negligible and corrections due  to the effects
that we have listed earlier, after the kinetic equations, are at most ${\cal O}(1)$ factors. On the other hand, as we will discuss in section \ref{sec:allowed_regions}, there are two interesting new favoured regions for leptogenesis around $z \pm \sim \pi/2$ for NH,
 where the decay asymmetry from $N_2$ decays dominates over the one from $N_1$ decays (`$N_2$-dominated regions').
Fig.~\ref{fig:K1alpha} shows that 
in this region $K_{2\a} \sim 2\div 5$. 
We are therefore
in a  `optimal washout' region where thermal leptogenesis works most efficiently but still the dependence on the initial conditions amounts not more
than $\sim 50\%$. We have therefore decided to show the results just for the case of vanishing initial $N_2$-abundance since
this is the most conservative case with lowest efficiency. Very similar results
are obtained for the other benchmark cases as well.

%We will see that in the end the small regions where
%the $K_{i\a}\ll 5$ and one can expect some dependence of the initial conditions, will
%mostly not coincide with the regions where leptogenesis is successful and will therefore not
%play a relevant role in the discussion.

\begin{figure}
\includegraphics [width=0.5\textwidth]{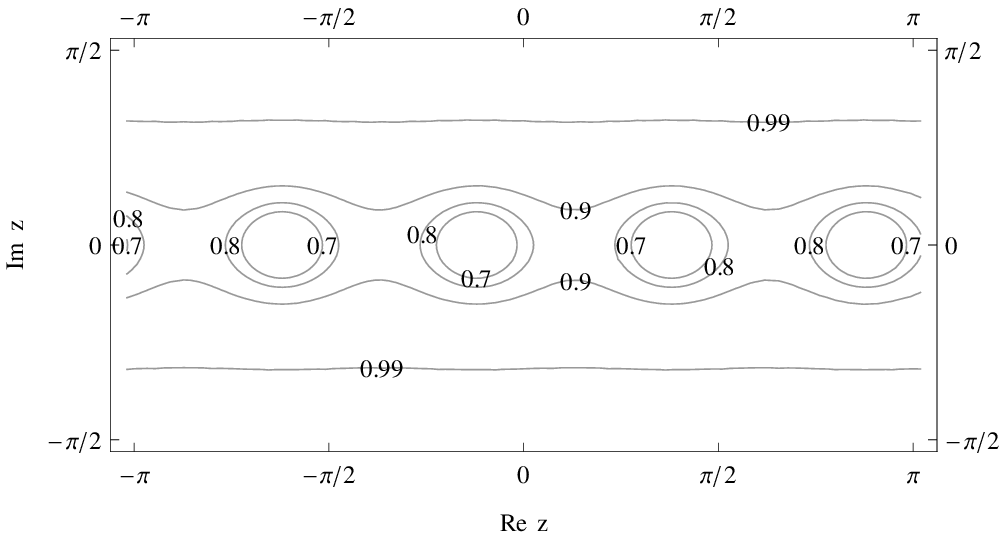}
\includegraphics [width=0.5\textwidth]{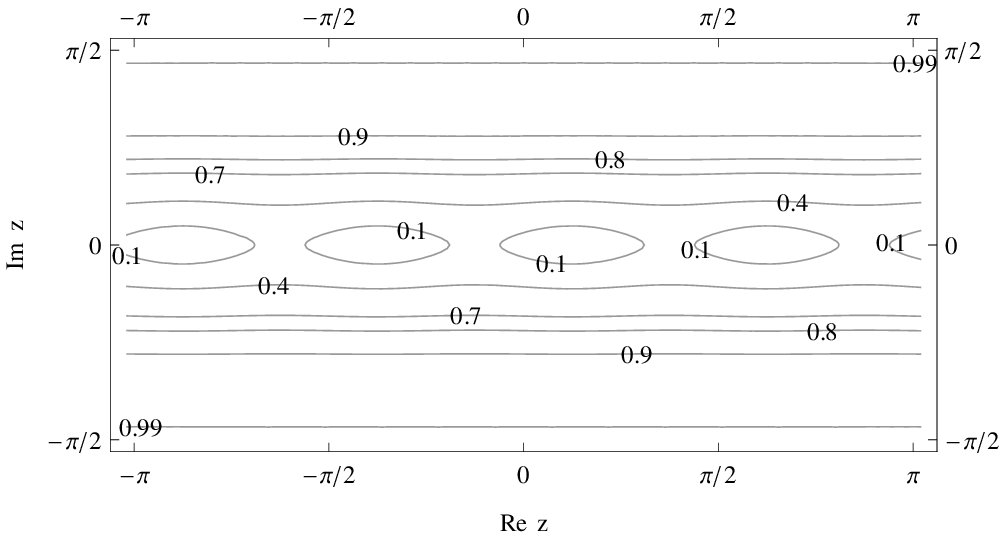}
 \vspace*{-7mm}
\caption{Contour plots of $p_{12}$ for NH (left) and IH (right), benchmark B, $\zeta = + 1$ .}
\label{fig:p12}
\end{figure}

At the same time, with $K_{1\g}$ and $K_{1\tau}\gg 1$, the asymmetries $\Delta_{\tau}$ and $\Delta_{\g_1}$
produced from $N_2$-decays are efficiently washed out by $N_1$ inverse processes, and practically only the orthogonal component $\Delta_{\g_1^{\bot}}$, with size determined by  $1 - p_{12}$, survives.
Fig.~\ref{fig:p12} shows the contour plot of $p_{12}$ which indicates that
the quantity significantly differs from unity in general. For NH $p_{12}$ is periodic in $\p$ along the $ {\rm Re}z$ axis and is approximately periodic in $\p /2$.
Notice also that for IH, $p_{12}$ depends on ${\rm Im}z$ only. 
One can already see that in the new favoured regions around $z \sim \pm \pi/2$ the quantity $1 - p_{12}$ is maximal. We thus find good conditions for leptogenesis regarding washout from $N_2$ as well as from $N_1$ processes.

\subsection{CP Asymmetries in the orthogonal parametrization}
%%%%%%%%%%%%%%%%%%%%%%%%%%%%%%%%%%%%%%%%%%%%%%%%%%%%%%%%%%%%%

Let us now re-express the $C\!P$ asymmetries in the orthogonal parametrization.
The expression (\ref{epsia}) for the $C\!P$ asymmetries can be recast as
\bea \label{eq:epsIal}
\varepsilon_{i \alpha}
        &\,=\,& -\frac{3}{16\,\pi v^2}
\, \frac{1}{(m_D^\dagger \, m_D)_{ii}}  \,
 \sum\limits_{j \neq i}  \, \left(
{\cal I}_{ij}^\alpha \, {\xi(M_j^2/M_i^2)\over M_j/M_i}
+ {\cal J}_{ij}^\alpha \, \frac{2}{3\,(M_j^2/M_i^2 -1)} \right)
\\ & \equiv & \ve^{\cal I}_{i\a} + \ve^{\cal J}_{i\a} \, ,
\eea
in an obvious notation where we have defined,
\bea \label{eq:calIJ}
{\cal I}_{ij}^\alpha \equiv
{\rm Im} \Big[ \big(m_D^\dagger \big)_{i \alpha} \, \big(m_D \big)_{\alpha j}
\big(m_D^\dagger m_D \big)_{ij} \Big]~,~~
{\cal J}_{ij}^\alpha \equiv {\rm Im}
\Big[\big (m_D^\dagger \big)_{i \alpha} \, \big(m_D \big)_{\alpha j} \big(m_D^\dagger m_D \big)_{ji} \Big] \,.
\eea
It is evident that ${\cal I}_{ij}^\alpha = - {\cal I}_{ji}^\alpha$
and ${\cal J}_{ij}^\alpha = - {\cal J}_{ji}^\alpha$.
In terms of the R-matrix we write from eq.~(\ref{R}),
\be
m_D=U\,D_{\sqrt{k}}\,R^T\, D_{\sqrt{M}}. \label{R3}
\ee
Then using this we find:
\be
{\cal I}_{ij}^\alpha =  M_i\, M_j \, {\rm Im} \Big[ \sum\limits_{k,k',k''}
(m_{k}m_{k'})^{1/2}m_{k''}U_{\alpha k}^* U_{\alpha k'}\, R^*_{ik}\,R_{jk'}R^*_{ik''}\,R_{jk''} \Big] \, ,
\ee
\be {\cal J}_{ij}^\alpha =  M_i \, M_j \, {\rm Im} \Big[ \sum\limits_{k,k',k''} (m_{k}m_{k'})^{1/2}m_{k''}U_{\alpha k}^*
U_{\alpha k'}\, R^*_{ik}\,R_{jk'}R^*_{jk''}\,R_{ik''} \Big] \, .
\ee

%We can now recast the different terms in the final asymmetry
%(cf. eq~{(\ref{NBmLf}), (\ref{NBmLf1}), (\ref{NBmLf2})) in the
%orthogonal parametrization. Let us then start from the
%lightest RH neutrino $C\!P$ asymmetries.

In order to simply the notation, it will prove convenient to introduce the ratios
\be
r_{i\alpha}\equiv {\varepsilon_{i\alpha}\over \bar{\ve}(M_1)} \, , \,\,\,
r^{\cal I}_{i\alpha}\equiv {\varepsilon^{\cal I}_{i\alpha}\over \bar{\ve}(M_1)} \, , \,\,\,
r^{\cal J}_{i\alpha}\equiv {\varepsilon^{\cal J}_{i\alpha}\over \bar{\ve}(M_1)} \, .
\ee
where
\be 
\bar{\ve}(M_1) \equiv {3\over 16\,\pi}\,{M_1\,m_{\rm atm}\over v^2} \simeq
10^{-6}\,\left({M_1\over 10^{10}\,{\rm GeV}}\right)\,
\ee
is the upper bound  for the total $C\!P$ asymmetries \cite{di}
that is therefore used as a reference value.

\subsubsection{Lightest RH neutrino $C\!P$ asymmetries}

We can start from the lightest RH neutrino $C\!P$ asymmetries $\ve_{1\a}$. We first
write them including the third heaviest RH neutrino corresponding to the terms $j=3$.
We need then to specialize the general expressions above for ${\cal I}_{ij}$ and
${\cal J}_{ij}$ to the case $i=1$ obtaining
\be
{\cal I}_{1j}^\alpha =  M_1\,M_j {\rm Im}
\Big[ \sum\limits_{k,k',k''} (m_{k}m_{k'})^{1/2}m_{k''}U_{\alpha k}^* U_{\alpha k'}\,
R^*_{1k}\,R_{jk'}R^*_{1k''}\,R_{jk''} \Big]
\ee
and
\be
{\cal J}_{1j}^\alpha =  M_1 M_j {\rm Im} \Big[
\sum\limits_{k,k',k''} (m_{k}m_{k'})^{1/2}m_{k''}U_{\alpha k}^* U_{\alpha k'}\,
R^*_{1k}\,R_{jk'}R^*_{jk''}\,R_{1k''} \Big] \, .
\ee
When we sum over $j$ in the first term of the eq.~(\ref{eq:epsIal}) for $i=1$ containing ${\cal I}_{1j}$,
thanks to $R$ orthogonality and considering that for
$M_2\gtrsim 3\,M_1$ we can approximate $\xi(M_j^2/M_1^2)\simeq 1$. Then,
only terms $k'=k''$ survive and one obtains \cite{abada2}
\be\label{r1aI}
 r_{1\alpha}^{\cal I}  = \sum_{k,k'}\, {m_{k'}\,\sqrt{m_{k'}\,m_k}\over \mt\,m_{\rm atm}} \,{\rm
Im}[U_{\a k}\,U_{\a k'}^{\star}\,R_{1k}\,R_{1k'}] \, ,
\ee
where the effective neutrino masses $\mti$ can be written in terms of the
R-matrix using eq.~(\ref{mtiR},).
This  term is bounded by \cite{flavoreffects,flavorlep}
\be\label{cpupperbound}
\left| r_{1\alpha}^{\cal I} \right|  < 
\sqrt{P^0_{1\alpha}}\,{{\rm max}_i[m_i]\over m_{\rm atm}}\,{\rm max}_k\,[|U_{\a k}|]  \\
\ee
and it is the only term that has been considered in all previous analyses of leptogenesis
in the two RH neutrino model so far.
It is  useful to give a derivation of this upper bound. One can first write
\bea
\left| r_{1\alpha}^{\cal I} \right|& \leq  & {1\over \mt\,m_{\rm atm}}\,
 \left|\sum_{k} \sqrt{m_k} \,U_{\a k}\,R_{1 k} \right| \,
 \left|\sum_{k'} (m_{k'})^{3\over 2} \,U^{\star}_{\a k'}\,R_{1 k'} \right|  \\
 & = & \sqrt{P^0_{1\alpha}}\,{{\rm max}_i[m_i]\over m_{\rm atm}} \, 
    \sqrt{\widetilde{P}_{1\a}^0} \, ,
 \eea
where in the second line we used the eq.~(\ref{P01a}) and defined the quantity
\be
\widetilde{P}_{1\a}^0 \equiv {\left|\sum_{k'} 
(m_{k'})^{3\over 2} \,U^{\star}_{\a k'}\,R_{1k'} \right|^2 \over ({\rm max}_i[m_i])^2 \, \mt} \, .
\ee
Considering the definition eq.~(\ref{mtiR}) for $\mt$, this can then be 
maximised writing
\be
\widetilde{P}_{1\a}^0 \leq  {\sum_{k'} 
m_{k'}\, |\,U^{\star}_{\a k'}\,R_{1k'}|^2 \over  \mt} \leq {\rm max}_k\,[|U_{\a k}|^2]  \,  .
\ee
In this way one obtains the upper bound eq.~(\ref{cpupperbound}).
For the other  term the situation is quite different.
The second term containing ${\cal J}_{1j}$ cannot be simplified
using the $R$ orthogonality and one obtains \cite{bounds}
\be \label{r1aJ}
r_{1\alpha}^{\cal J} =
-{2\over 3}\,\sum_{j,k,k',k''} {M_1\over M_j}\,\, \,{m_{k''}\sqrt{m_{k}\,m_{k'}}\over \mt\,m_{\rm atm}}\,
{\rm Im}[U^{\star}_{\a k}\,U_{\a
k'}\,R^{\star}_{1k}\,R_{jk'}\,R^{\star}_{jk''}\,R_{1k''}] \, .
\ee
Let us now specialize the expressions eqs.~(\ref{r1aI}) and (\ref{r1aJ}) for $r_{1\alpha}^{\cal I}$
and $r_{1\alpha}^{\cal J}$ to the two RH neutrino case
using the  special forms for the orthogonal matrix R
(cf. (\ref{RNH}) and (\ref{RIH})) for NH and IH respectively.
One can immediately check that the $j=3$ terms vanish and for NH one obtains \cite{abada2}
\bea\label{r1alphaI}
r_{1\a}^{\cal I} & = & {m_{\rm atm}\over \mt}\,{\rm Im}[\sin^2 z]\,
\left(|U_{\a 3}|^2 - {m_{\rm sol}^2\over m_{\rm atm}^2}\,|U_{\a 2}|^2 \right) \\ \nonumber
& + &  \zeta\,{\sqrt{m_{\rm sol}\,m_{\rm atm}}\over \mt\,m_{\rm atm}}\,\left\{(m_{\rm atm}-m_{\rm sol})\,
{\rm Im}[U_{\a2}\,U^{\star}_{\a 3}]\,{\rm Re}[\sin z\,\cos z] \right. \\ \nonumber
& + & \left. (m_{\rm atm}+m_{\rm sol})\,{\rm Re}[U_{\a2}\,U^{\star}_{\a 3}]\,{\rm Im}[\sin z\,\cos z] \right\}
\eea
and
\bea\label{r1alphaJ}
-r_{1\a}^{\cal J} & = & {2\over 3}\,{M_1\over M_2}\left\{\,{m_{\rm sol}\over \mt}\,
{\rm Im}[\sin^2 z]\,\left(|U_{\a 3}|^2-|U_{\a 2}|^2\right)\right. \\ \nonumber
& + & \zeta\,{\sqrt{m_{\rm atm}\,m_{\rm sol}}\over \mt\,m_{\rm atm}}\,\left[(m_{\rm atm}-m_{\rm sol})\,
{\rm Im}[U^{\star}_{\a 2}\,U_{\a 3}]\,{\rm Re}[\sin z\,\cos^{\star} z]\,
(|\cos z|^2+|\sin z|^2)\right. \\ \nonumber
& + & \left.\left. (m_{\rm atm}+m_{\rm sol})\,
{\rm Re}[U^{\star}_{\a 2}\,U_{\a 3}]\,{\rm Im}[\sin z\,\cos^{\star} z]\,
(|\cos z|^2 -|\sin z|^2) \right]\right\}
\, .
\eea
In terms of ${\rm Re}z , \, {\rm Im}z$ the dominant term $r_{1\a}^{\cal I}$ is
\bea\label{r1alphaIz}
r_{1\a}^{\cal I} & = & {m_{\rm atm}\over \mt}\, \frac{1}{2}\sin[2{\rm Re}z]\sinh[2{\rm Im}z]\,
\left(|U_{\a 3}|^2 - {m_{\rm sol}^2\over m_{\rm atm}^2}\,|U_{\a 2}|^2 \right) \\ \nonumber
& + & \frac{1}{2}\,\zeta\,{\sqrt{m_{\rm sol}\,m_{\rm atm}}\over \mt\,m_{\rm atm}}\,\left\{(m_{\rm atm}-m_{\rm sol})\,
{\rm Im}[U_{\a2}\,U^{\star}_{\a 3}]\,\sin[2{\rm Im}z]\,\cosh[2{\rm Im}z] \right. \\ \nonumber
& + & \left. (m_{\rm atm}+m_{\rm sol})\,{\rm Re}[U_{\a2}\,U^{\star}_{\a 3}]\,\cos[2{\rm Re}z]\sinh[2{\rm Im}z] \right\} \, .
\eea
%This expression is particularly helpful in understanding the dependence of a final $N_1$-dominated asymmetry on $z$ and $U$.

\begin{figure}
 \includegraphics [width=0.5\textwidth]{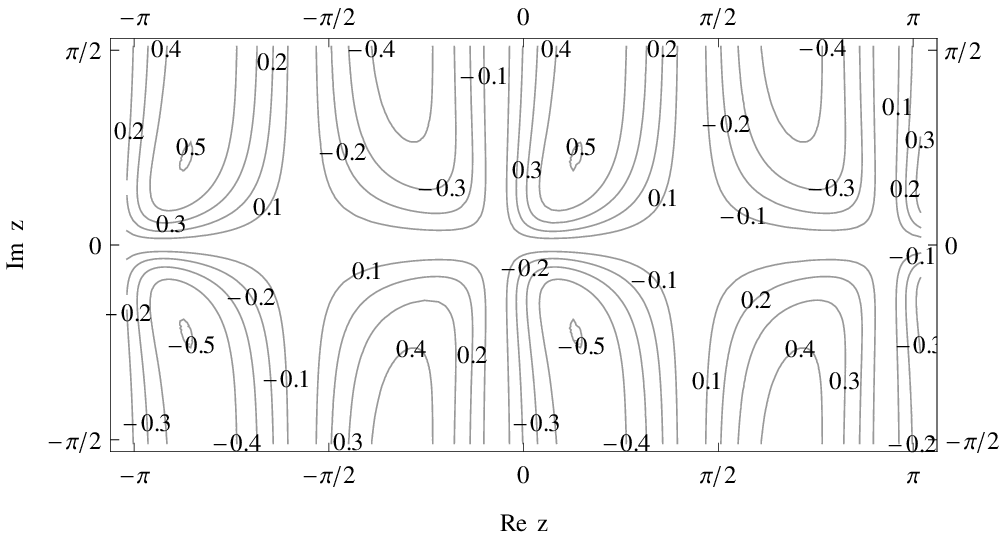}
 \includegraphics[width=0.5\textwidth]{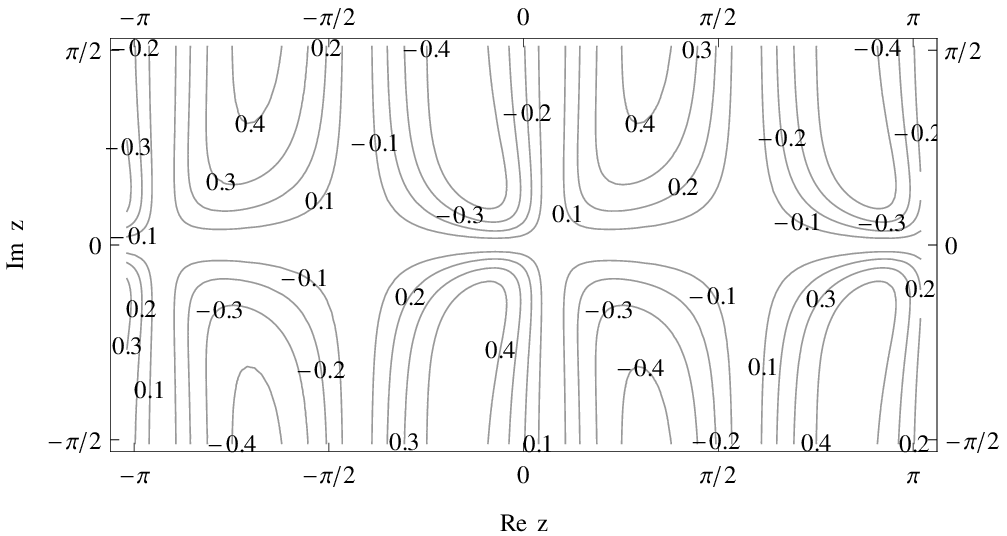}
 \includegraphics[width=0.5\textwidth]{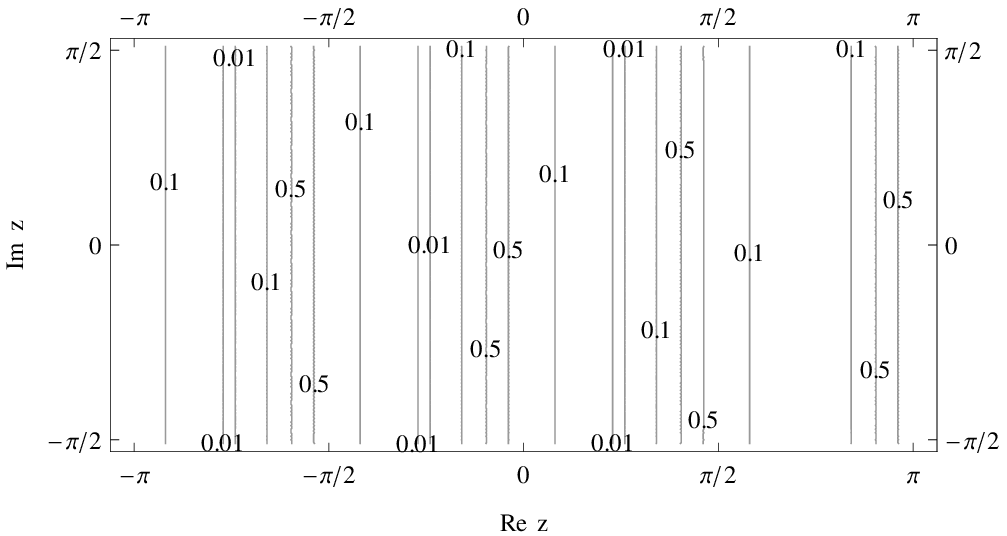}
 \includegraphics[width=0.5\textwidth]{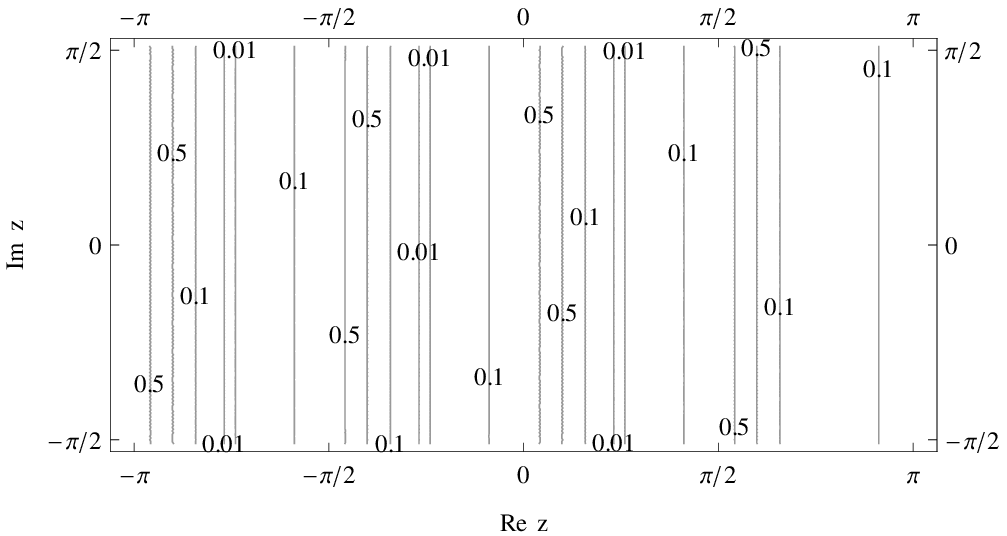}
\vspace*{-7mm}
\caption{Contour plots of $r^{\cal I}_{1\g}$ (upper left panel) , $r^{\cal I}_{1\t}$ (upper right panel), $|r^{\cal J}_{1\g}/r^{\cal I}_{1\g}|= |\ve^{\cal J}_{1\g}/\ve^{\cal I}_{1\g}|$ (lower left panel) and  $|r^{\cal J}_{1\t} /r^{\cal I}_{1\t}|= |\ve^{\cal J}_{1\t}/\ve^{\cal I}_{1\t}|$ (lower right panel) for NH, benchmark B eq.~(\ref{bench1}), $\zeta=+1$ and $M_2/M_1=3$.}
\label{r1alpha}
\end{figure}

Analogously, for the case of IH and approximating $m_1\simeq m_2 \simeq m_{\rm atm}$, one obtains
\be
r_{1\a}^{\cal I}  =
{m_{\rm atm}\over \mt}\,
\left\{{\rm Im}[\sin^2 z]\,(|U_{\a 1}|^2\,-|U_{\a 2}|^2)
- 2\, \zeta\,{\rm Re}[U_{\a1}\,U^{\star}_{\a 2}]\,{\rm Im}[\sin z\,\cos z]
\right\}
\ee
and
\bea
r_{1\a}^{\cal J}  &\simeq &  {2\over 3}\,{M_1\over M_2}\,{m_{\rm atm}\over \mt}\,
\left\{{\rm Im}[\sin^2 z]\,(|U_{\a 1}|^2 - |U_{\a 2}|^2) \right. \\ \nonumber
& + & \left. 2\, \zeta\,\left(|\sin z|^2-|\cos z|^2 \right)\,{\rm Re}[U_{\a1}\,U^{\star}_{\a 2}]\,{\rm Im}[\sin z\,\cos^{\star} z]   \right\} \, .
\eea
Notice that while the terms $r_{1\a}^{\cal J}$ are proportional to $M_1/M_2$, the terms  $r_{1\a}^{\cal I}$ are not.
In Figure~\ref{r1alpha} we have plotted the quantities $r^{\cal I}_{1\a}$ and
$|r^{\cal J}_{1\a}/r^{\cal I}_{1\a}| = |\ve^{\cal J}_{1\a}/\ve^{\cal I}_{1\a}|$ for the benchmark B $U_{PMNS}$ choice eq.~(\ref{bench2}),
$\zeta=+1$ and  $M_2/M_1=3$. Once again there is periodicity in $\p$ along ${\rm Re}z$, for the same reasons as with Figs.~\ref{fig:K1alpha},\ref{fig:p12}.
\begin{figure}
 \includegraphics[width=0.5\textwidth]{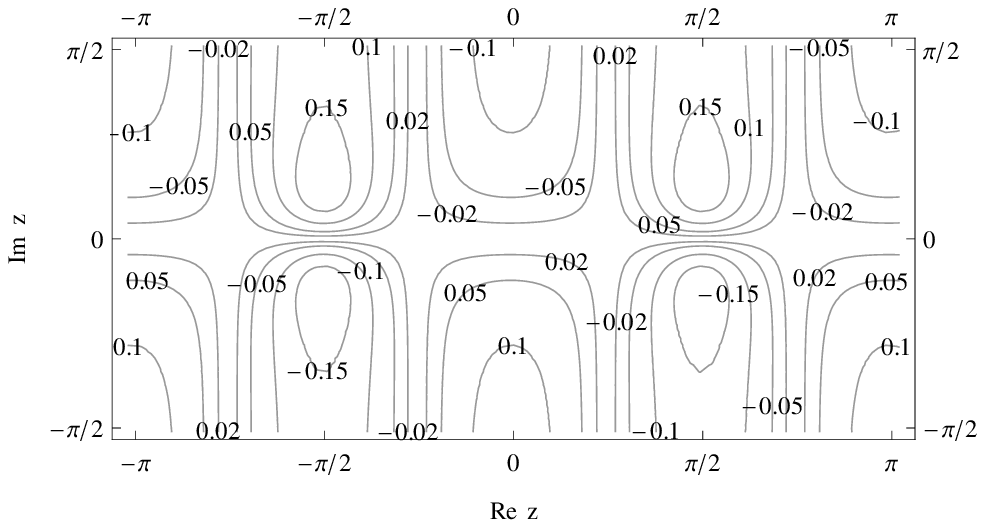}
 \includegraphics[width=0.5\textwidth]{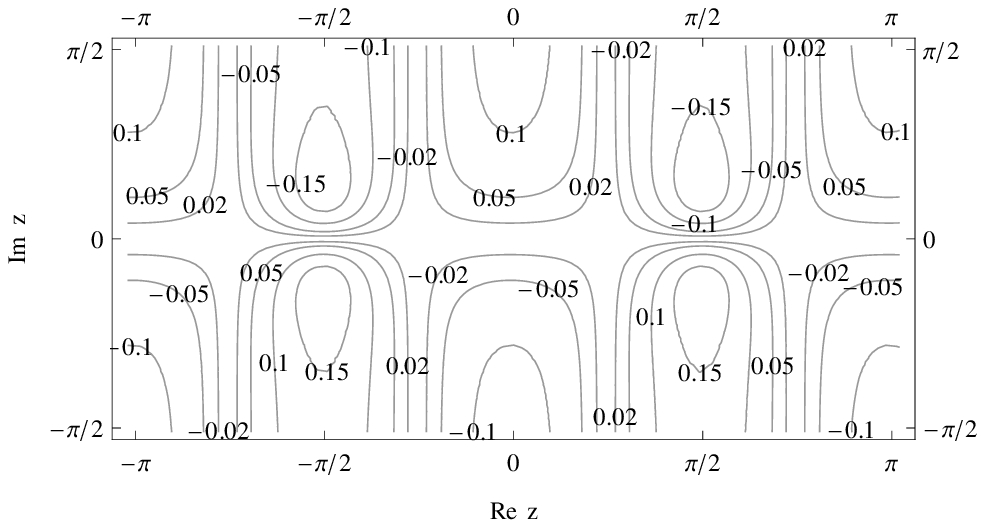}
 \includegraphics[width=0.5\textwidth]{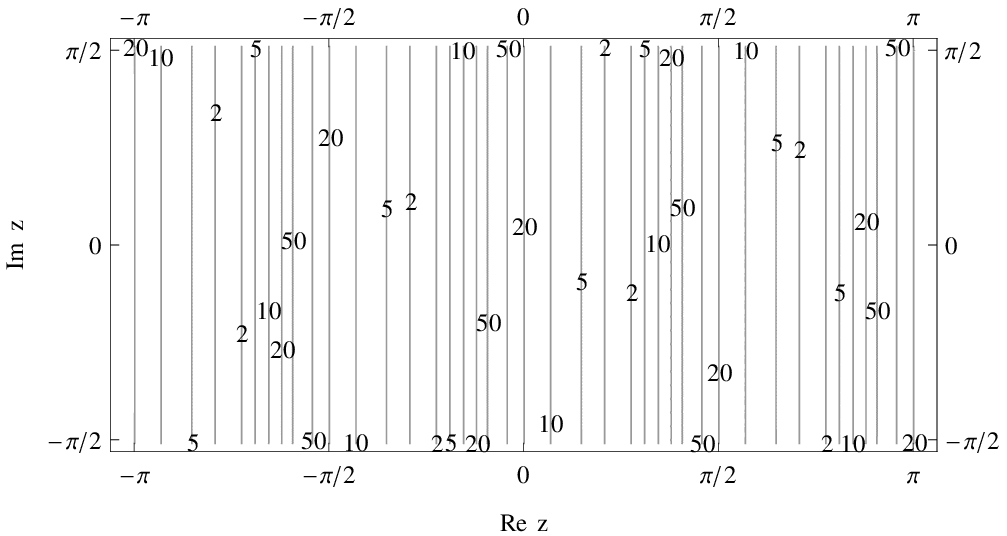}
 \includegraphics[width=0.5\textwidth]{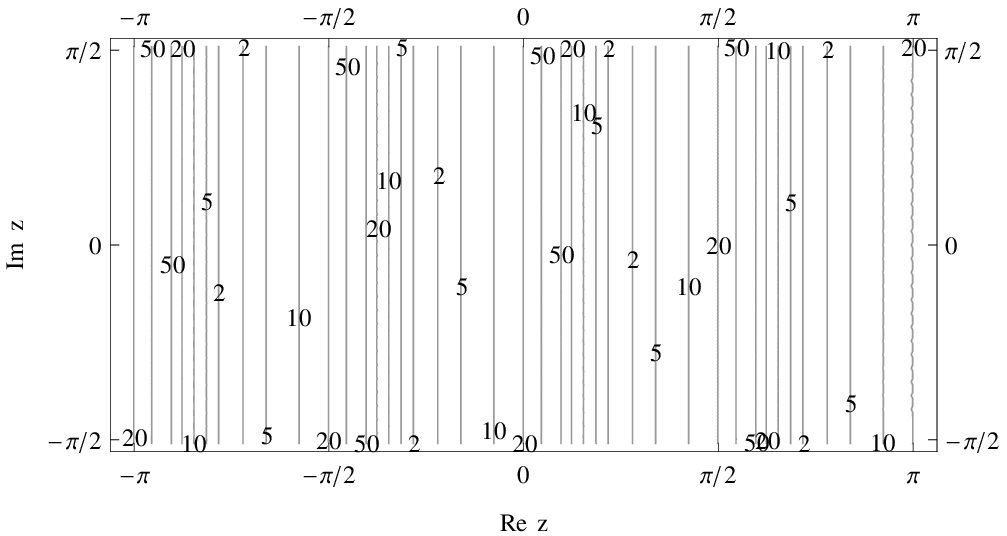}
\vspace*{-7mm}
\caption{Contour plots of $r^{\cal J}_{2\g}$ (upper left panel), $r^{\cal J}_{2\t}$ (upper right panel),
$|r^{\cal J}_{2\g}/r^{\cal I}_{2\g}|=|\epsilon^{\cal J}_{2\g}/\epsilon^{\cal I}_{2\g}|$ (lower left panel) and
$|r^{\cal J}_{2\t} /r^{\cal I}_{2\t}|=|\ve^{\cal J}_{2\t} /\ve^{\cal I}_{2\t}|$
(lower right panel) for benchmark B, eq.~(\ref{bench2}), $\zeta=+1$, $M_2/M_1=3$ and NH.
}
\label{r2alpha}
\end{figure}
One can notice how $|r^{\cal J}_{1\a}/r^{\cal I}_{1\a}|\ll 1$ both for $\a=\g$ and $\a=\t$. Only in a very
fine tuned region this ratio gets up to about $0.5$.
Therefore, it will prove out the term $r^{\cal J}_{1\a}$ can be safely neglected in the regions of interest for this study.
Also, one can notice that $r^{\cal I}_{1\a}$, the dominant contribution to the baryon asymmetry from $N_1$ decays, is suppressed in the region $z\sim \p /2$, hence this region is potentially dominated by $N_2$ decays.

 From the lower panels of fig.~\ref{r1alpha} one can notice how $|r^{\cal J}_{1\a} /r^{\cal I}_{1\a}|$ is independent of ${\rm Im}[z]$, eq.~(\ref{bench2}). This is because ${\rm Im}[U_{\a 2}\,U^{\star}_{\a 3}]=0$ such that the middle terms vanish from eqs.~(\ref{r1alphaI}) and (\ref{r1alphaJ}). It can then be shown that all the dependance of $r_{1\a}^{\cal I}$ and $r_{1\a}^{\cal J}$ on ${\rm Im}[z]$ is contained in the common factor $\sinh[2{\rm Im}z] \, / \mt$ which then cancels in the ratio $|r^{\cal J}_{1\a} /r^{\cal I}_{1\a}|$.

\subsubsection{Next-to-lightest RH neutrino}

Let us now turn to consider the case $i=2$. For $j=3$ we have
\be
{\cal I}_{23}^\alpha =
M_2\,M_3\,\sum_{k,k',k''} m_{k''}\,\sqrt{m_k\,m_{k'}}\, {\rm Im}[U_{\alpha k}^* U_{\alpha
k'}\,R^*_{2k}\,R_{3k'}\,R^*_{2k''}\,R_{3k''}] \, ,
\ee
and
\be {\cal J}_{23}^\alpha = M_2\,M_3\,\sum_{k,k',k''}\, m_{k''}\,\sqrt{m_{k'}\,m_{k}}\, {\rm Im}[U_{\alpha k}^* U_{\alpha
k'}\,R^*_{2k}\,R_{3k'}\,R^*_{3k''}\,R_{2k''}] \, .
\ee
It is easy to check that both two terms vanish in the two RH neutrino case.

On the other hand the two terms for $j=1$,
\be {\cal I}_{21}^\alpha = M_2\,M_1\,\sum_{k,k',k''} m_{k''}\,\sqrt{m_k\,m_{k'}}\, {\rm Im}[U_{\alpha k}^* U_{\alpha
k'}\,R^*_{2k}\,R_{1k'}\,R^*_{2k''}\,R_{1k''}] \, , \ee
and
\be {\cal J}_{21}^\alpha = M_2\,M_1\,\sum_{k,k',k''}\,
m_{k''}\,\sqrt{m_{k'}\,m_{k}}\, {\rm Im}[U_{\alpha k}^* U_{\alpha k'}\,R^*_{2k}\,R_{1k'}\,R^*_{1k''}\,R_{2k''}] \, ,
\ee
do not vanish and they lead, in the hierarchical limit $M_2\gtrsim 3\,M_1$,
to final values for $r_{2\alpha}^{\cal I}$
and $r_{2\alpha}^{\cal J}$ given respectively by
\be
r_{2\alpha}^{\cal I} \simeq  -{4 \over 3} \, \left({M_1 \over M_2}\right)
\left[\ln  \left({M_2 \over M_1}\right) - 1 \right]
\, \sum_{k,k',k''} {m_{k''}\,\sqrt{m_k\,m_{k'}} \over
\tilde{m}_2 \, m_{atm} }\, {\rm Im}[U_{\alpha k}^* U_{\alpha k'}\,R^*_{2k}\,R_{1k'}\,R^*_{2k''}\,R_{1k''}] \, ,
\ee
and
\be
r_{2\alpha}^{\cal J} \simeq
{2\over 3}\,\sum_{k,k',k''}\, {m_{k''}\,\sqrt{m_{k'}\,m_{k}} \over \tilde{m}_2 \, m_{atm} }\,
{\rm Im}[U_{\alpha k}^*\, U_{\alpha k'}\,R^*_{2k}\,R_{1k'}\,R^*_{1k''}\,R_{2k''}] \, .
\ee
This second term $r_{2\alpha}^{\cal J}$ will clearly tend to dominate over
$r_{2\alpha}^{\cal I} \propto (M_1/M_2)$. However, since the dependence on the
complex parameter $z$ is different, one cannot exclude that, in some region of the parameter
space, $r_{2\alpha}^{\cal I}$ can give a non negligible contribution. We have therefore safely
taken into account this term checking indeed that is negligible.

If we specialize the expressions to the two RH neutrino model we obtain for NH

\bea
r_{2\a}^{\cal I} & = & -{4\over 3}\,{M_1\over M_2}\,\left[\ln\left({M_2\over M_1}\right)-1\right]\,
\left\{\,{m_{\rm atm}\over \mt}\,
{\rm Im}[\sin^2 z]\,\left[|U_{\a 3}|^2-{m_{\rm sol}^2\over m_{\rm atm}^2}\,|U_{\a 2}|^2\right]\right. \\ \nonumber
& + &  \zeta\,{\sqrt{m_{\rm atm}\,m_{\rm sol}}\over \mtt\,m_{\rm atm}}\,(m_{\rm atm}-m_{\rm sol})\,
{\rm Im}[U_{\a 2}\,U^{\star}_{\a 3}]\,{\rm Re}[\sin z\,\cos^{\star} z]\,
[|\cos z|^2+|\sin z|^2] \\ \nonumber
& + & \left.  \zeta\,{\sqrt{m_{\rm atm}\,m_{\rm sol}}\over \mtt\,m_{\rm atm}}\,(m_{\rm atm}+m_{\rm sol})\,
{\rm Re}[U_{\a 2}\,U^{\star}_{\a 3}]\,{\rm Im}[\sin z\,\cos^{\star} z]\,
[|\cos z|^2 -|\sin z|^2] \right\}
\eea
and
\bea
r_{2\a}^{\cal J} & = & {2\over 3}\,{m_{\rm sol}\over \mtt}\,{\rm Im}[\sin^2 z]\,\left[|U_{\a 2}|^2-|U_{\a 3}|^2 \right] \\ \nonumber
& + & {2\over 3}\, \zeta\,{\sqrt{m_{\rm atm}\,m_{\rm sol}}\over \mtt \,m_{\rm atm}}\,(m_{\rm atm}-m_{\rm sol})\,
{\rm Im}[U^{\star}_{\a 2}\,U_{\a 3}]\,{\rm Re}[\sin z\,\cos^{\star} z]\,
[|\cos z|^2+|\sin z|^2] \\ \nonumber
& + & {2\over 3}\,\zeta\,{\sqrt{m_{\rm atm}\,m_{\rm sol}}\over \mtt \,m_{\rm atm}}\,(m_{\rm atm}+m_{\rm sol})\,
{\rm Re}[U^{\star}_{\a 2}\,U_{\a 3}]\,{\rm Im}[\sin z\,\cos^{\star} z]\,
[|\cos z|^2 -|\sin z|^2]
\, ,
\eea
In terms of ${\rm Re}z , \, {\rm Im}z$ the dominant term $r_{2\a}^{\cal J}$ is
\bea\label{r2alphaJz}
r_{2\a}^{\cal J} & = & {1\over 3}\,{m_{\rm sol}\over \mtt}\,\sin[2{\rm Re}z]\sinh[2{\rm Im}z]\,\left[|U_{\a 2}|^2-|U_{\a 3}|^2 \right] \\ \nonumber
& + & {1\over 3}\, \zeta\,{\sqrt{m_{\rm atm}\,m_{\rm sol}}\over \mtt \,m_{\rm atm}}\,(m_{\rm atm}-m_{\rm sol})\,
{\rm Im}[U^{\star}_{\a 2}\,U_{\a 3}]\,\sin[2{\rm Re}z]\cosh[2{\rm Im}z]\,
\\ \nonumber
& + & {1\over 3}\,\zeta\,{\sqrt{m_{\rm atm}\,m_{\rm sol}}\over \mtt \,m_{\rm atm}}\,(m_{\rm atm}+m_{\rm sol})\,
{\rm Re}[U^{\star}_{\a 2}\,U_{\a 3}]\,\cos[2{\rm Re}z]\sinh[2{\rm Im}z]
\, ,
\eea
%The above expression are particularly helpful in understanding the dependence of a final $N_2$ dominated asymmetry on $z$ and $U$.
For IH we obtain
\bea
r_{2\a}^{\cal I} & = & {m_{\rm atm}\over \mt}\,
\left\{{\rm Im}[\sin^2 z]\,\left(|U_{\a 1}|^2-|U_{\a 2}|^2 \right) \right. \\ \nonumber
& + & 2\, \zeta\,\left(|\sin z|^2\,{\rm Im}[U^{\star}_{\a 1}\,U_{\a 2}]\,{\rm Re}[\sin z\,\cos^{\star} z]\right. \\ \nonumber
& + & \left.  \;\;\;\; |\cos z|^2\,{\rm Re}[U^{\star}_{\a 1}\,U_{\a 2}]\,{\rm Im}[\sin z\,\cos^{\star} z]  \right)
\eea
and
\bea
r_{2\a}^{\cal J} & = &
{m_{\rm atm}\over \mtt}\,
\left\{{\rm Im}[\sin^2 z]\,\left(|U_{\a 1}|^2 -|U_{\a 2}|^2 \right)\, \right. \\ \nonumber
& + & \left. 2\, \zeta\,{\rm Re}[U_{\a1}\,U^{\star}_{\a 2}]\,{\rm Im}[\sin z\,\cos^{\star} z]\,
[|\sin z|^2 -|\cos z|^2] \right\} \, .
\eea
In Figure~\ref{r2alpha} we have plotted $r^{\cal I}_{2\a}$ and
$|r^{\cal J}_{2\a}/r^{\cal I}_{2\a}|$ for $\zeta=+1$, $M_2/M_1=3$,
benchmark $U_{PMNS}$ choice B (c.f.\ eq.~(\ref{bench2})) and NH.
This time, as one can see from the figures, one has $|r^{\cal J}_{2\a}/r^{\cal I}_{2\a}|\gg 1$ for all values of $z$ and $M_1/M_2$
(since $|r^{\cal J}_{2\a}/r^{\cal I}_{2\a}|$ gets even larger if $M_2/M_1 > 3$ ), implying that the
term $r^{\cal J}_{2\a}$ dominates and that $r^{\cal I}_{2\a}$ can be safely neglected. It can again be seen in fig.~\ref{r2alpha}  that
$r^{\cal J}_{2\a}/r^{\cal I}_{2\a}$ depends only on ${\rm Re }z$  for the same reasons as with $|r^{\cal J}_{1\a}/r^{\cal I}_{1\a}|$. Once again there is periodicity in $\p$ along ${\rm Re}z$, for the same reasons as with Figs.~\ref{fig:K1alpha},\ref{fig:p12},\ref{r1alpha}.

Crucially, we find that $r^{\cal J}_{2\a}$, the dominant contribution to the baryon asymmetry from $N_2$ decays, is maximised in the regions $z\sim \pm \p / 2$ (just above and below the $ {\rm Im} z = 0$ line), in contrast to $r^{\cal I}_{1\a}$, the dominant contribution from $N_1$ decays, which is minimised in this region (see fig~\ref{r1alpha}). Given this result and the favourable values of $K_{2\g}$ and $p_{12}$, shown in fig~\ref{fig:K1alpha} and fig~\ref{fig:p12} respectively, one expect the $z \sim \pm \p / 2$ regions will be $N_2$ dominated.

Notice that we have not shown any figure for the case of IH since it will turn out that the
contribution from the next-to-lightest RH neutrinos to the final asymmetry
is always negligible.

%%%%%%%%%%%%%%%%%%%%%%%%%%%%%%%%%%%%%%%%%%%%%%%%%%%%%%%%%%%%%%%%%%%%%%%%%%%%%%%%%%%
\section{Constraints on the parameter space and $N_1$ versus $N_2$ contribution}\label{sec:allowed_regions}
%%%%%%%%%%%%%%%%%%%%%%%%%%%%%%%%%%%%%%%%%%%%%%%%%%%%%%%%%%%%%%%%%%%%%%%%%%%%%%%%%%%

We can now finally go back to the expression for the final asymmetry (cf. eqs.(\ref{NBmLf}), (\ref{NBmLf1}) and (\ref{NBmLf2}))
and recast it within the orthogonal parametrization. This can be written as the sum of four terms,
\be\label{finalas}
N_{B-L}^{\rm f}=N_{B-L}^{\rm f (1,{\cal I})}+N_{B-L}^{\rm f (1,{\cal J})}+N_{B-L}^{\rm f (2,{\cal I})}+N_{B-L}^{\rm f (2,{\cal J})} \, .
\ee
The sum of the  first two terms is the contribution $N_{B-L}^{\rm f (1)}$ from the lightest RH neutrinos,
\bea
N_{B-L}^{\rm f (1,{\cal I})}(z,U,M_1)& =  & - \bar{\ve}(M_1)\,\sum_{\a=\t,\g} \,r_{1\a}^{\cal I}(z,U)\,\k_{1\a} \, ,\\
N_{B-L}^{\rm f (1,{\cal J})}(z,U,M_1,M_1/M_2) & =  & - \bar{\ve}(M_1)\,\sum_{\a=\t,\g} \,r_{1\a}^{\cal J}(z,U,M_1/M_2)\,\k_{1\a} \, ,
\eea
and it should be noticed that only the second one depends on $M_2$.

Analogously the sum of the last two terms  in eq.~(\ref{finalas}) is the
contribution $N_{B-L}^{\rm f (2)}$ from the next-to-lightest RH neutrinos,
\bea
N_{B-L}^{\rm f (2,{\cal I})}(z,U,M_1,M_1/M_2) & = & - \bar{\ve}(M_1)\,{M_1\over M_2}\,\left[\ln\left({M_2\over M_1}\right)-1\right]\,f^{\cal I}(z,U) \, , \\
N_{B-L}^{\rm f (2,{\cal J})}(z,U,M_1) & = & - \bar{\ve}(M_1)\,f^{\cal J}(z,U)\, ,
\eea
where
\bea
{M_1\over M_2}\,\,\left[\ln\left({M_2\over M_1}\right)-1\right]\,f^{\cal I}(z,U) & = & p_{12}\,r^{\cal I}_{2\gamma}\,\k_{2\g}\,e^{-{3\,\pi\,\over 8}K_{1\gamma} }
\\ \nonumber
& + & (1-p_{12})\,r^{\cal I}_{2\gamma}\,\k_{2\g} + 
r^{\cal I}_{2\tau}\,\k_{2\tau}\,e^{-{3\,\pi\,\over 8}K_{1\tau} }
\eea
and
\be
f^{\cal J}(z,U)=p_{12}\,r^{\cal J}_{2\gamma}\,\k_{2\g}\,
e^{-{3\,\pi\,\over 8}K_{1\gamma} } +
(1-p_{12})\,r^{\cal J}_{2\gamma}\,\k_{2\g} + 
r^{\cal J}_{2\tau}\,\k_{2\tau}\,e^{-{3\,\pi\,\over 8}\,K_{1\tau} } \, .
\ee
Notice that this time  the first term depends on $M_2$ while the second does not.
We can then write the total asymmetry as
\be
N_{B-L}^{\rm f}= [N_{B-L}^{\rm f (1,{\cal I})}+N_{B-L}^{\rm f (2,{\cal J})}](z,U,M_1)\,
[1+\delta_1 + \delta_2](z,U,M_1,M_1/M_2) \, ,
\ee
where we defined $\delta_1\equiv N_{B-L}^{\rm f (1,{\cal J})}/[N_{B-L}^{\rm f (1,{\cal I})}+N_{B-L}^{\rm f (2,{\cal J})}]$
and $\delta_2\equiv N_{B-L}^{\rm f (2,{\cal I})}/[N_{B-L}^{\rm f (1,{\cal I})}+N_{B-L}^{\rm f (2,{\cal J})}]$.
We found that $\delta_1,\delta_2\lesssim 0.05$ for any choice of $M_1/M_2,z,U$. Therefore, one can conclude
that the total final asymmetry is independent of $M_2$ with very good accuracy
\footnote{Notice that all $C\!P$ asymmetries, and consequently the final asymmetry, are $\propto M_1$.
Therefore, there will be still a lower bound on $M_1$ contrarily to the 3 RH neutrino
scenarios considered in \cite{N2dominated,us}.}
. 
It should be however
remembered that our calculation of $N_{B-L}^{\rm f (2)}$ holds for $M_2 \lesssim 10^{12}\,{\rm GeV}$ and $M_2/M_1\gtrsim 3$,
implying $M_1 \lesssim ( 100 \, / \, 3) \times 10^{10}\,{\rm GeV}$. As such, when $N_2$ decays are included the largest value of $M_1$ we will allow is $M_1 = 30 \times 10^{10}{\rm GeV}$, whereas when $N_2$ decays are neglected, we will consider values as large as $M_1 = 100 \times 10^{10}{\rm GeV}$.

\begin{figure}
 \includegraphics[width=.5\textwidth]{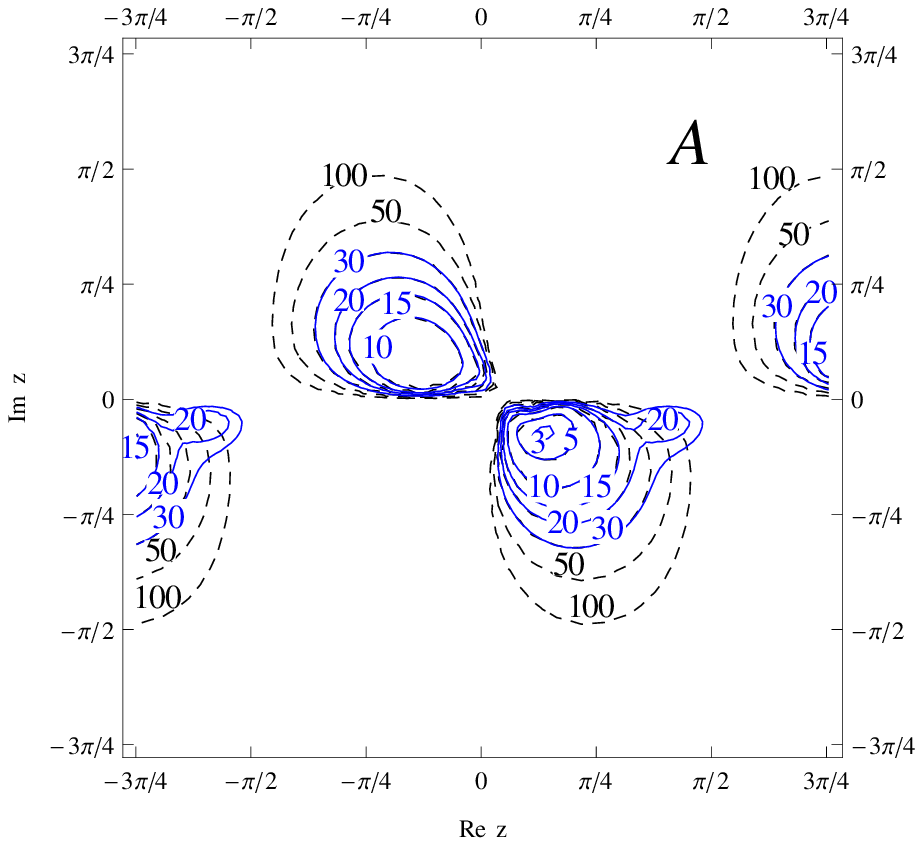}
 \includegraphics[width=.5\textwidth]{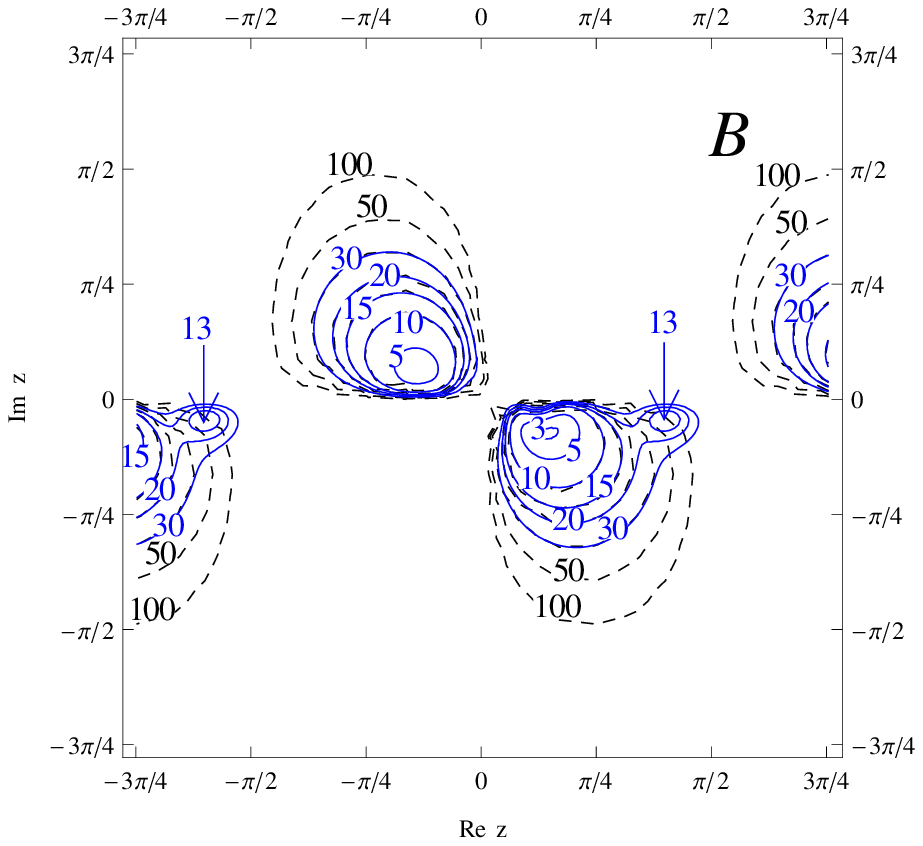}
 \includegraphics[width=.5\textwidth]{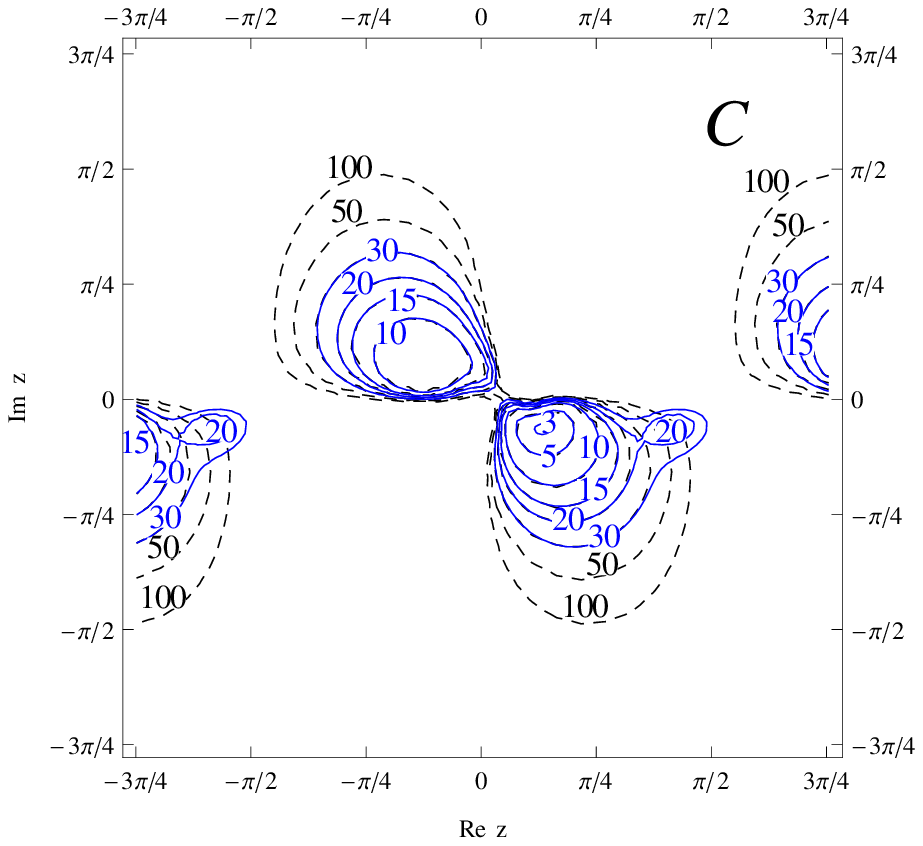}
 \includegraphics[width=.5\textwidth]{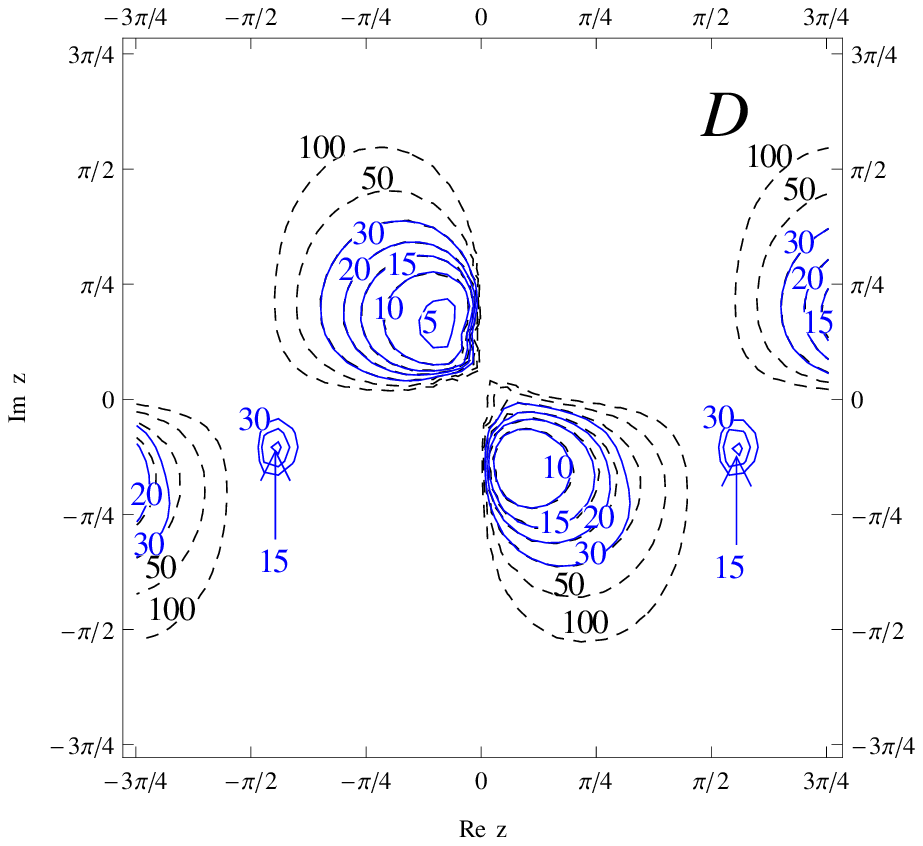}
\caption{Contours plots in the $z$-plane of the $M_1$ values obtained imposing successful leptogenesis ($\eta_B=\eta_B^{CMB}$)
for the NH case, $\zeta=+1$ and benchmarks A (top left), B (top right), C (bottom left) and D (bottom right) fixing U.
The solid lines are obtained taking into account the contribution $N_{B-L}^{\rm f (2)}$ to the final asymmetry
while the dashed lines are obtained neglecting this contribution.
  Contours are labelled with the value of $M_1$ in units of $10^{10}$GeV.}
\label{NHlbM1}
\end{figure}

In Fig.~\ref{NHlbM1} we show the contours plots for  $M_1$ obtained imposing successful
leptogenesis, i.e.  $\eta=\eta_B^{\rm CMB}$ (we used the $2\sigma$ lower value $\eta_B^{CMB}=5.9\times 10^{-10}$), for $\zeta=+1$
and for initial thermal abundance.
The four panels correspond to the four benchmark cases $A$, $B$, $C$ and $D$ in the NH case.
The solid lines are obtained including the contribution  $N_{B-L}^{\rm f (2)}$ in the final asymmetry
and therefore represent the main result of the paper. These have
to be compared with  the dashed lines where this contribution is neglected. In Fig.~\ref{NHlbM1} and indeed in all subsequent figures, one can notice again a periodicity in  $\p$ along ${\rm Re}z$. This is because final asymmetries are given from eq.~(\ref{finalas}), for which all terms are dependant upon quantities periodic in $\p$ along ${\rm Re}z$ (these quantities being the washouts, $p_{12}$ and the CP asymmetries).
As one can see, on most of the regions leptogenesis is $N_1$-dominated as one would expect
\footnote{The $N_1$-dominated regions are approximately invariant for $z\ra -z$,
implying $N_{B-L}^{\rm f (1,{\cal I})}(z)\simeq N_{B-L}^{\rm f (1,{\cal I})}(-z)$.
This is because $r_{1\a}^{\cal I}$ is dominated by the first term in the eq.~(\ref{r1alphaIz}),
exactly invariant for $z\ra -z$, and because the $K_{1\a}$ are also approximately invariant for $z\ra -z$ (see upper panels in Fig. 1).}.
However, there are two regions, around $z\sim \pm\pi/2$, where the asymmetry is $N_2$-dominated. If
$N_{B-L}^{\rm f (2)}$ is neglected, this region would be only partially accessible and in any case
only for quite large values  $M_1\gtrsim 30\times 10^{10}\,{\rm GeV}$
\footnote{Notice that this time there is no invariance with respect to $z\ra -z$ since
$r_{2\a}^{\cal J}$ is dominated either by the third term (for cases A, B, C,) or by the second term (for case D)
in eq.~(\ref{r2alphaJz}) that are not invariant for $z\ra -z$. On the other hand the
second term is invariant for  ${\rm Re}\,z \ra -{\rm Re}\,z$ and therefore one could naively expect a specular region at $z \sim -\pi/2$.
However, notice that $K_{2\gamma}$ is not invariant for  ${\rm Re}\,z\ra -{\rm Re}\,z$. In this way, for negative ${\rm Re}\,z$
and same values of $|z|$,  the wash-out is strong and prevents the existence of this specular region.}.

When the contribution $N_{B-L}^{\rm f (2)}$ is taken into account one can have
successful leptogenesis for $M_1$ values as low as $1.3\times 10^{11}\,{\rm GeV}$
for benchmark case B and vanishing initial $N_2$-abundance.
The existence of these `$N_2$-dominated regions' is the result of a combination
of different effects: i) the value of $(1-p_{12})$, setting the size of the contribution from
$N_2$ decays that survives the $N_1$ washout, is maximal in these regions as one can see from Fig.~\ref{fig:p12}; ii)
 the wash-out at the production is in these region minimum
as one can see from the plots of $K_{2\t}$ and $K_{2\g}$ (cf. Fig.~\ref{fig:K1alpha});
iii) the $N_2$-flavoured $C\!P$ asymmetries  are not suppressed in these regions
contrarily to the $N_1$ flavoured $C\!P$ asymmetries.
It is interesting to compare the results obtained for the 4 different benchmark cases. A comparison between
the case A (upper left panel) and the case $B$ (upper right panel) shows that  large values of $\theta_{13}$
and no Dirac phase enhance $N_{B-L}^{\rm f (2)}$ so that the $N_2$-dominated regions get enlarged. On the other hand
a comparison between B and C shows that a Dirac phase seems to suppress $N_{B-L}^{\rm f (2)}$. A comparison
between B and D shows that a Majorana phase seems just to change the position of the $N_2$-dominated regions without
consistently modify their size differently from the $N_1$-dominated regions that are instead
maximised by the presence of non-vanishing Majorana phase as known \cite{flavorlep,mp}.
Notice that,  though this effect is shown only for $\theta_{13}=11.5^{\circ}$, 
it actually occurs for any choice of $\theta_{13}$, in particular for $\theta_{13}=0$. 
Interestingly, for non-zero Majorana phase the new region where leptogenesis is favoured now overlaps with the Im$(z)=0$ axis. This means that CP violation
for $N_2$-dominated  leptogenesis can be successfully induced just by the Majorana phase.
We have also checked that varying the low energy parameters within the ranges of values
set by the 4 benchmark cases, one has a continuous variation of the allowed regions.

\begin{figure}
 \includegraphics[width=1\textwidth]{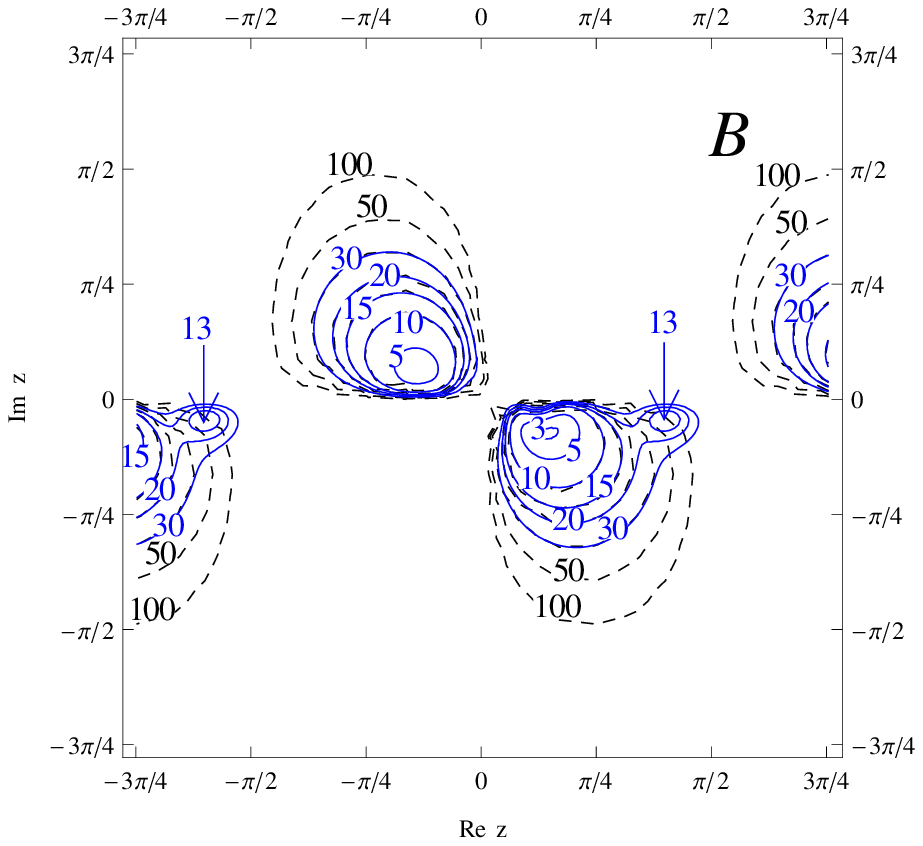}
\caption{Contours plots in the $z$-plane of the $M_1$ values obtained imposing successful leptogenesis ($\eta_B=\eta_B^{CMB}$)
for the NH case, $\zeta=+1$. Enlargement of the benchmark B case from previous figure. This is a particularly interesting case,
since it maximises the $N_2$ dominated  region around $z \approx z_{LSD} = \p/2$.}
\label{enlargement}
\end{figure}

It can be seen that the $N_2$-dominated regions are maximal in case B. For this reason in Fig.~\ref{enlargement}
we show a zoom of the $N_2$-dominated regions around $z=\pi/2$ for case B. This figure represents
one of the main results of this paper.

\begin{figure}
 \includegraphics[width=.5\textwidth]{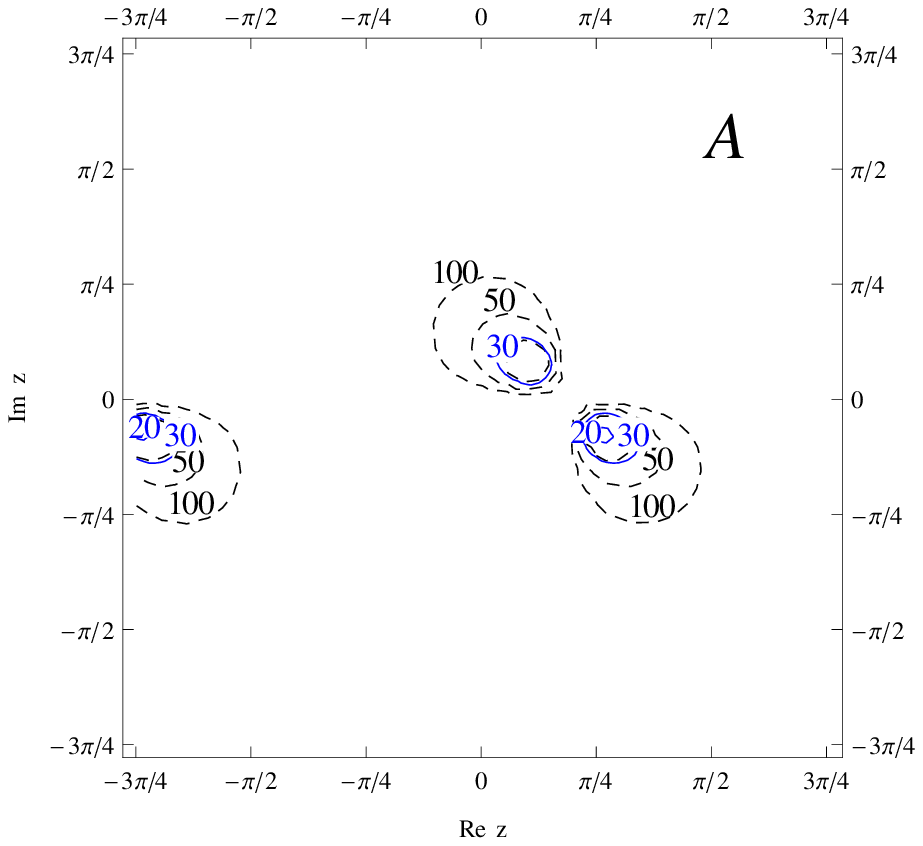}
 \includegraphics[width=.5\textwidth]{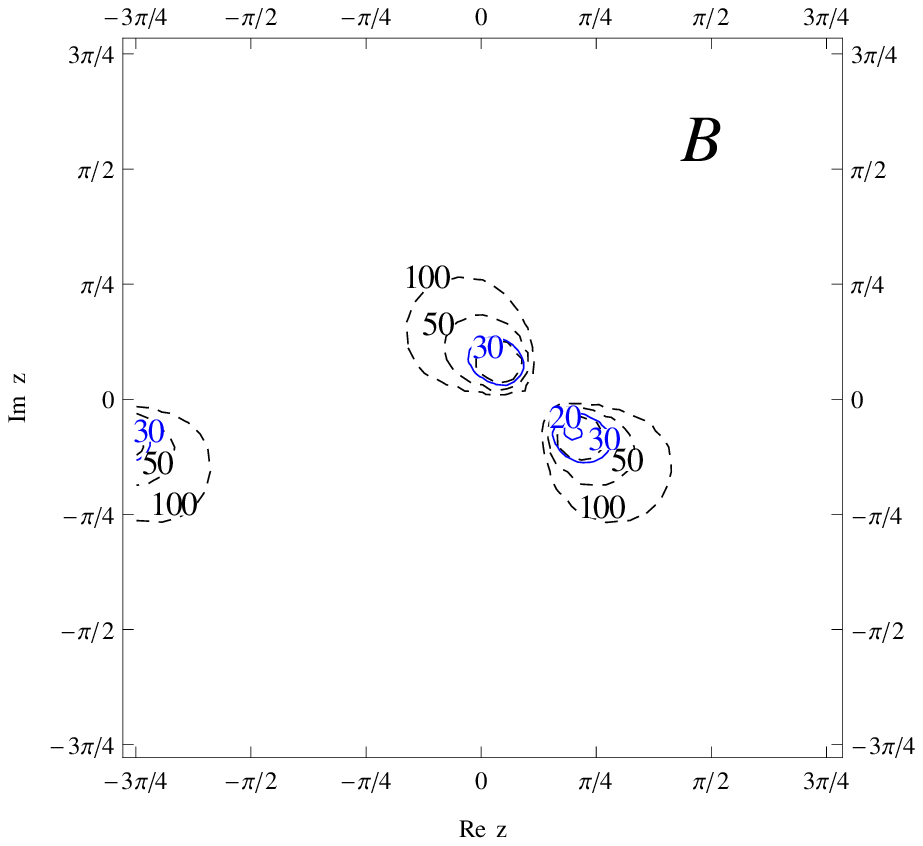}
 \includegraphics[width=.5\textwidth]{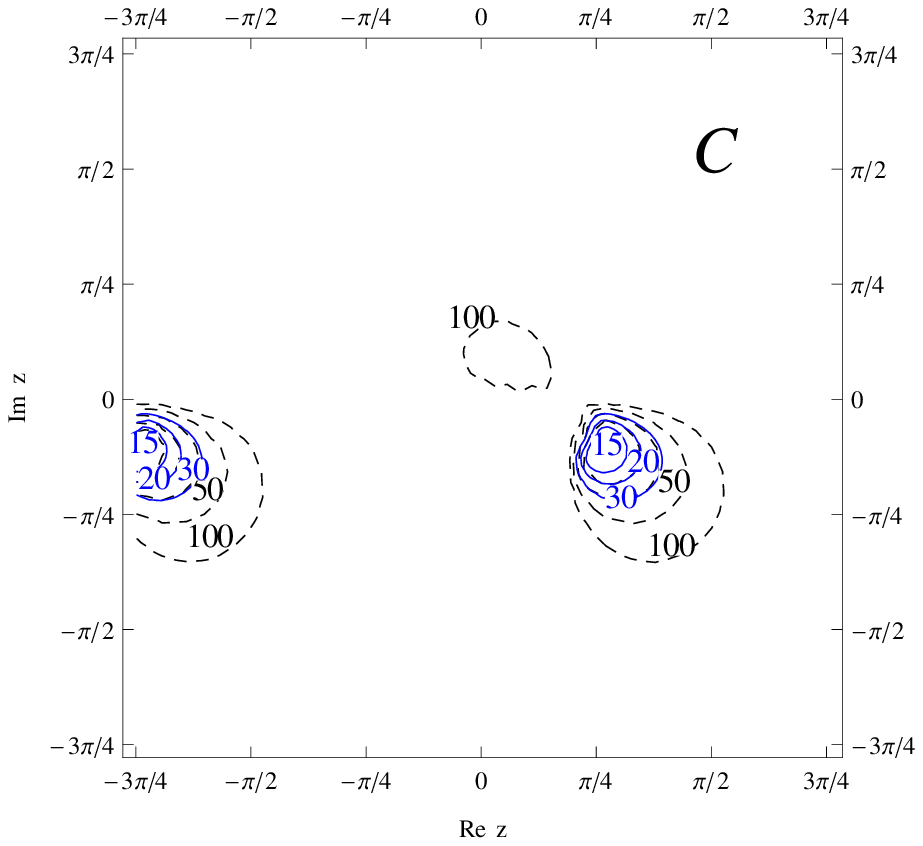}
 \includegraphics[width=.5\textwidth]{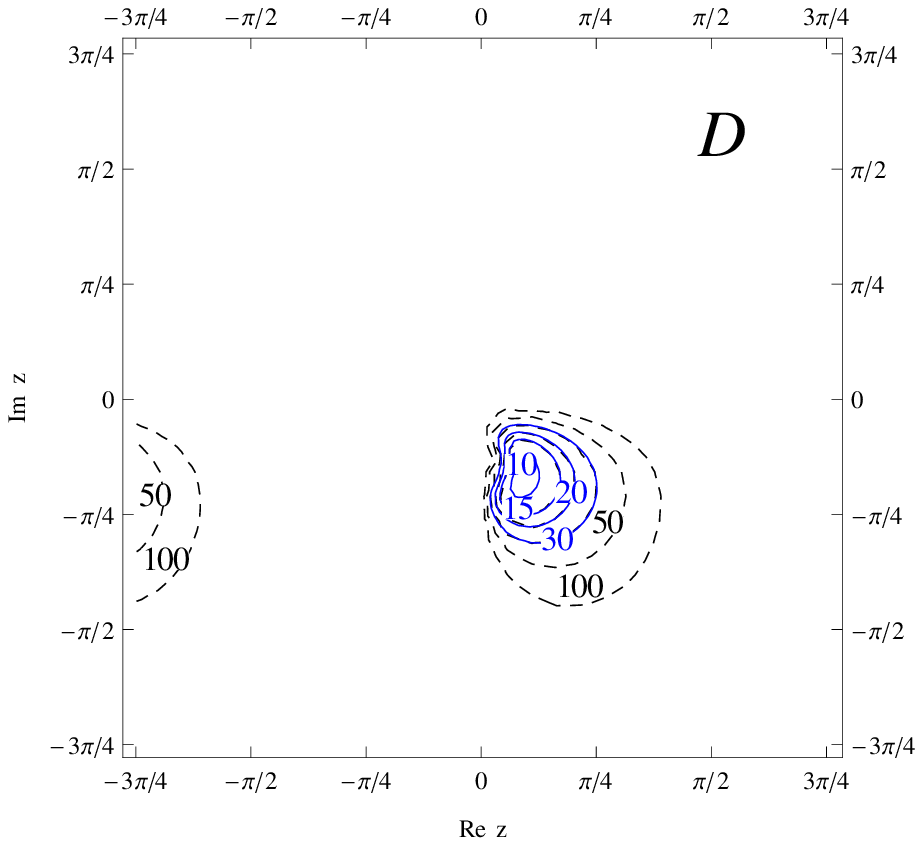}
\caption{Contours plots in the $z$-plane of the $M_1$ values obtained imposing successful leptogenesis ($\eta_B=\eta_B^{CMB}$)
for the IH case, $\zeta=+1$ and benchmarks A (top left), B (top right), C (bottom left) and D (bottom right) fixing U.
The solid lines are obtained taking into account the contribution $N_{B-L}^{\rm f (2)}$ to the final asymmetry
while the dashed lines are obtained neglecting this contribution.
  Contours are labelled with the value of $M_1$ in units of $10^{10}\,$GeV.}
\label{IHlbM1}
\end{figure}

\begin{figure}
 \includegraphics[width=.5\textwidth]{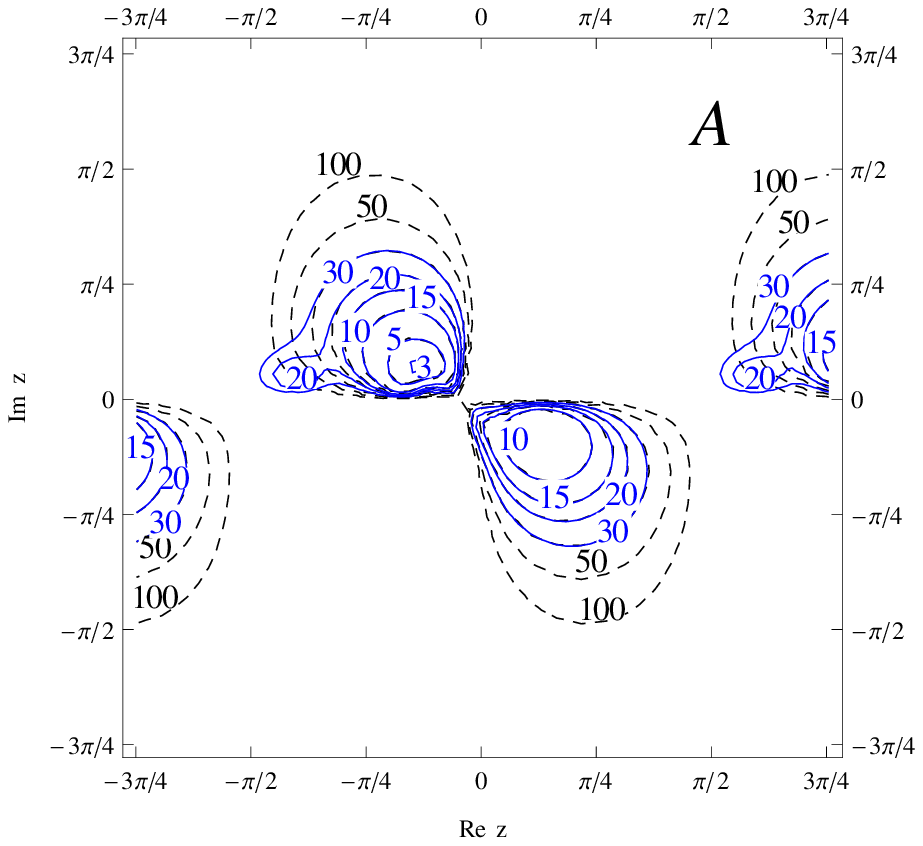}
 \includegraphics[width=.5\textwidth]{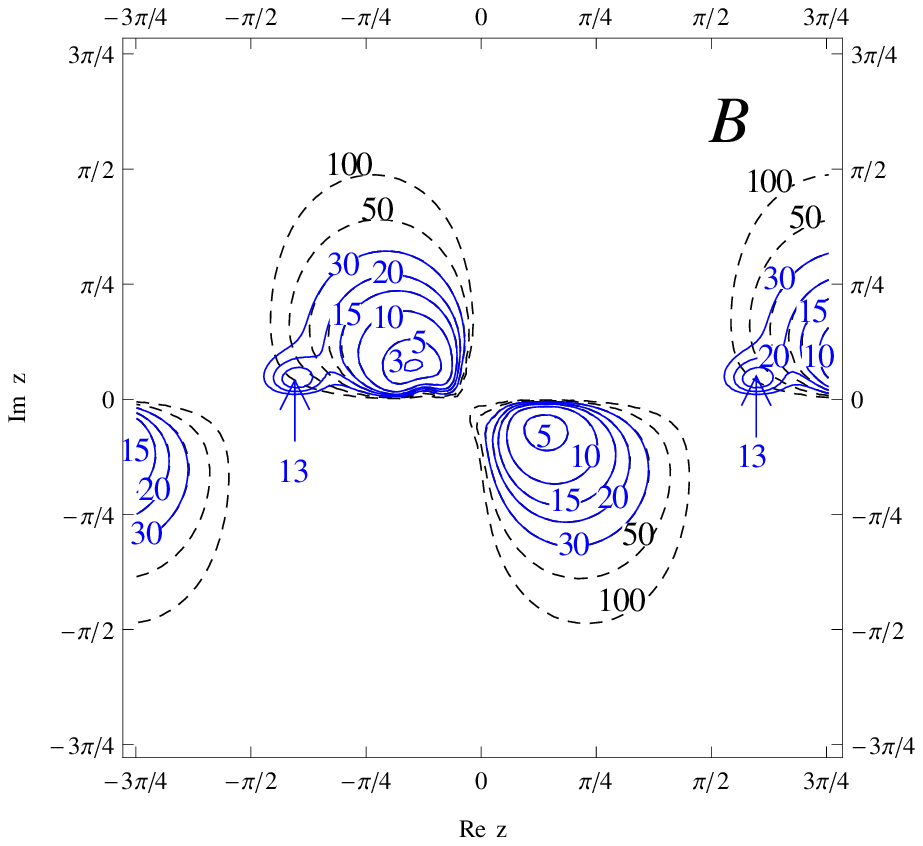}
 \includegraphics[width=.5\textwidth]{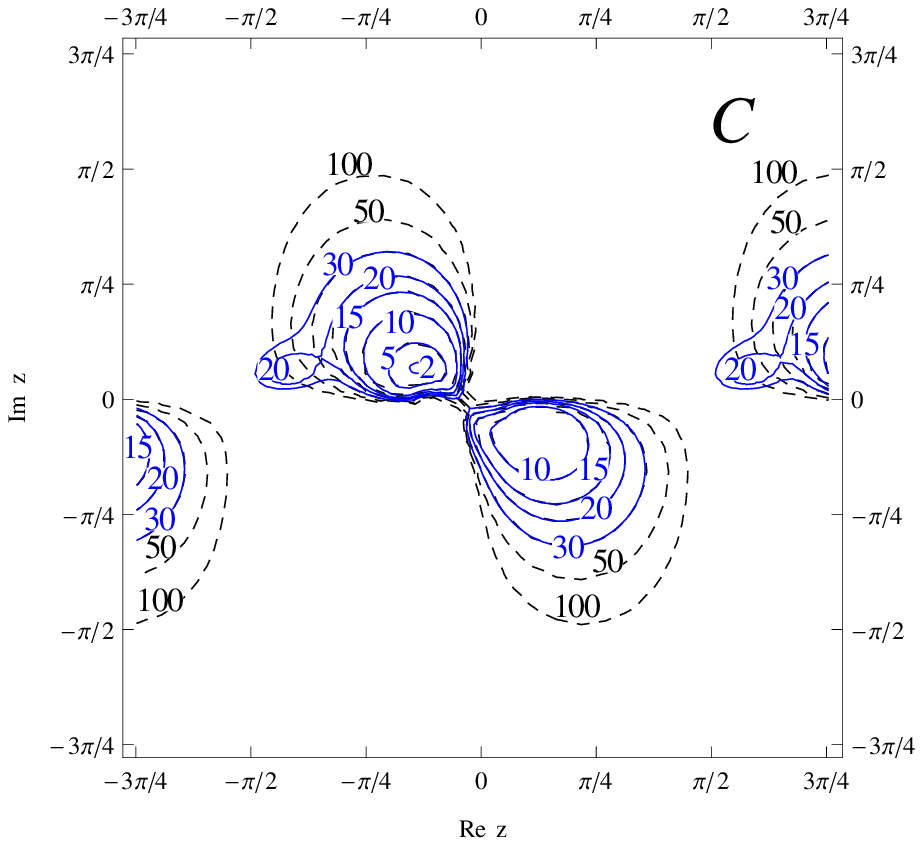}
 \includegraphics[width=.5\textwidth]{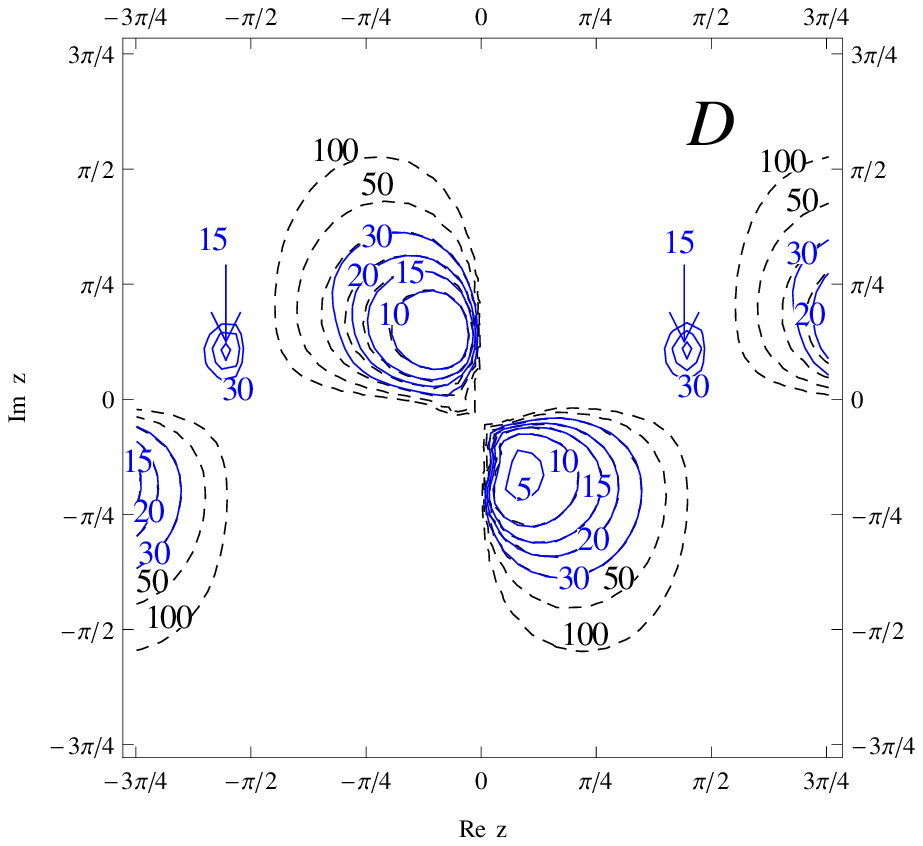}
\caption{Contours plots in the $z$-plane of the $M_1$ values obtained imposing successful leptogenesis ($\eta_B=\eta_B^{CMB}$)
for the NH case, $\zeta=-1$ and benchmarks A (top left), B (top right), C (bottom left) and D (bottom right) fixing U.
The solid lines are obtained taking into account the contribution $N_{B-L}^{\rm f (2)}$ to the final asymmetry
while the dashed lines are obtained neglecting this contribution.
  Contours are labelled with the value of $M_1$ in units of $10^{10}$GeV.}
\label{NHlbM1-}
\end{figure}

On the other hand if we consider the IH case, the situation is very different
as one can see from Fig.~\ref{IHlbM1}. The  much stronger wash-out acting both on the $N_1$
and on the $N_2$ contributions suppresses the final asymmetry in a way that
large fraction of the allowed regions disappear, including the $N_2$-dominated regions.
The surviving allowed regions are therefore strongly reduced and strictly $N_1$-dominated.
Analogous results are obtained for the branch $\zeta=-1$, shown
in figure \ref{NHlbM1-}.
\begin{figure}
 \includegraphics[width=1\textwidth]{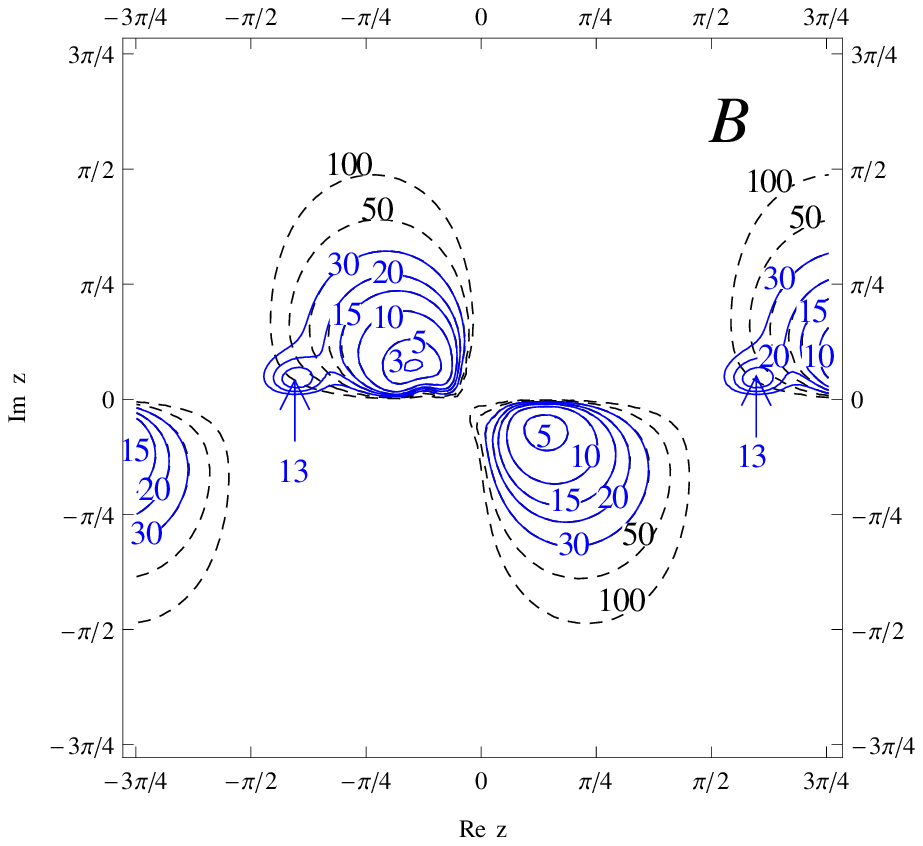}
\caption{Contours plots in the $z$-plane of the $M_1$ values obtained imposing successful leptogenesis ($\eta_B=\eta_B^{CMB}$)
for the NH case, $\zeta=-1$. Enlargement of the benchmark B case from the previous figure. This is a particularly interesting case,
since it maximises the $N_2$ dominated  region around $z \approx z_{LSD} = -\p/2$.}
\label{enlargement-}
\end{figure}
\begin{figure}
 \includegraphics[width=.5\textwidth]{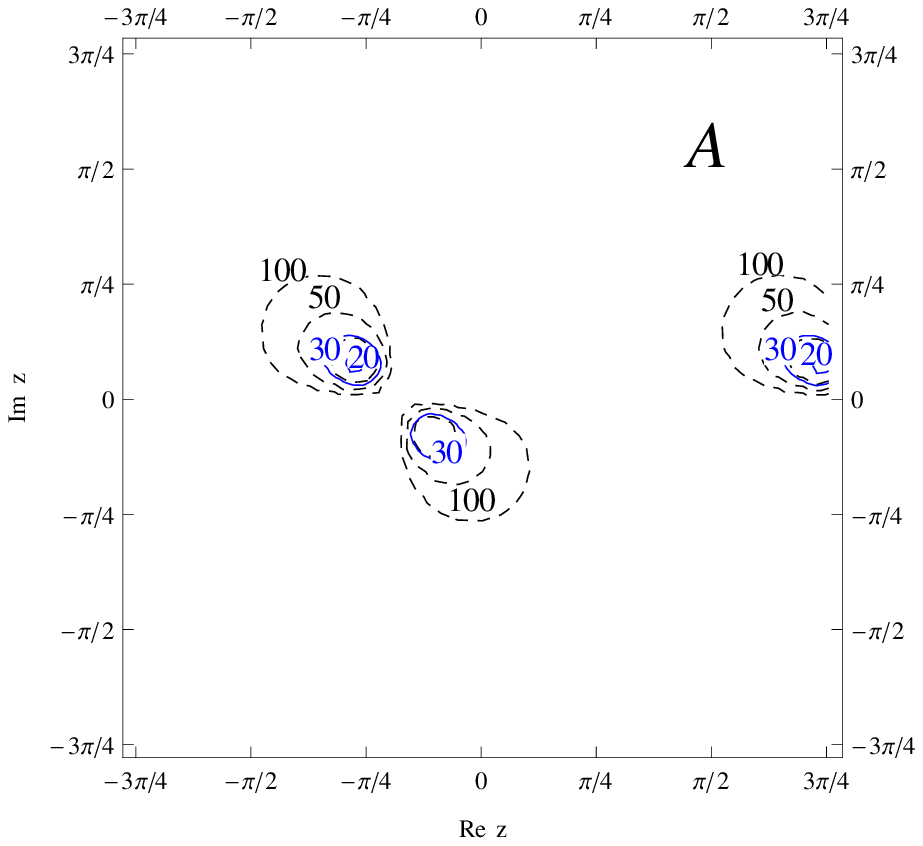}
 \includegraphics[width=.5\textwidth]{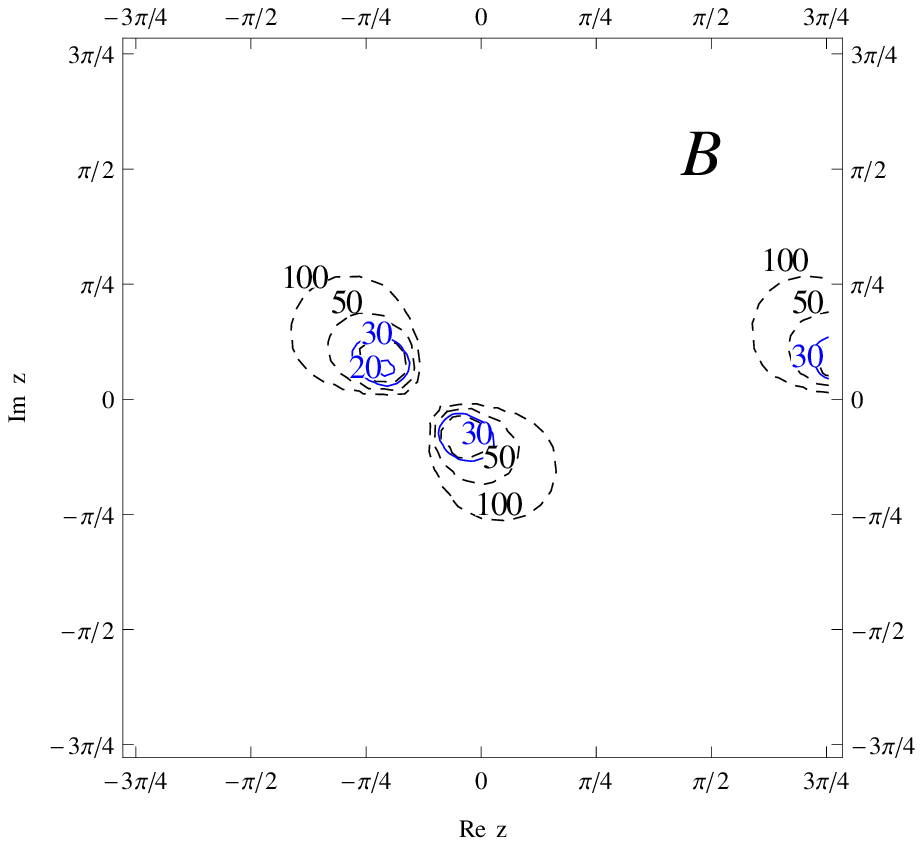}
 \includegraphics[width=.5\textwidth]{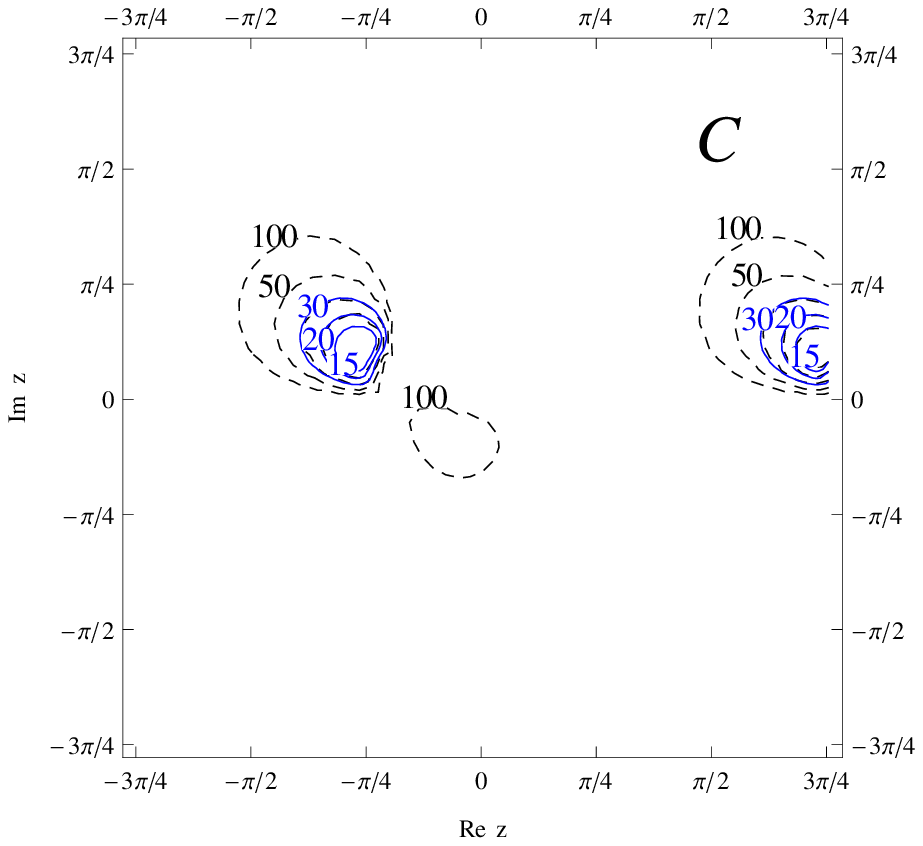}
 \includegraphics[width=.5\textwidth]{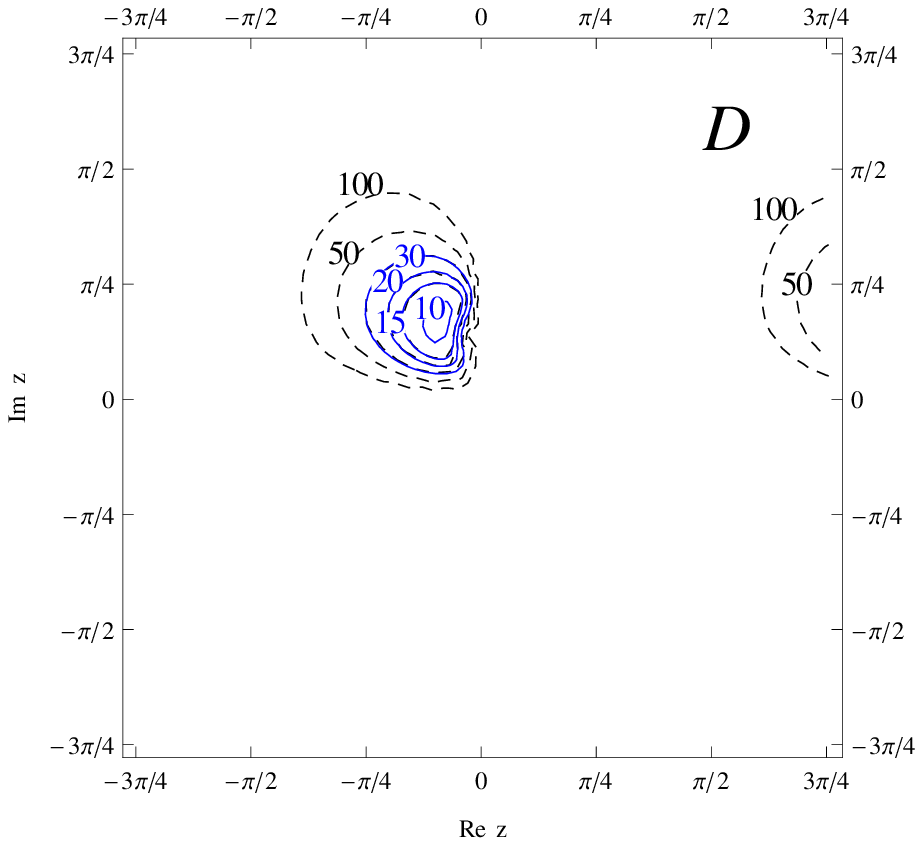}
\caption{Contours plots in the $z$-plane of the $M_1$ values obtained imposing successful leptogenesis ($\eta_B=\eta_B^{CMB}$)
for the IH case, $\zeta=-1$ and benchmarks A (top left), B (top right), C (bottom left) and D (bottom right) fixing U.
The solid lines are obtained taking into account the contribution $N_{B-L}^{\rm f (2)}$ to the final asymmetry
while the dashed lines are obtained neglecting this contribution.
  Contours are labelled with the value of $M_1$ in units of $10^{10}$GeV.}
\label{IHlbM1-}
\end{figure}
A comparison between the plots obtained for the two branches shows that the
the finally asymmetry is invariant for $(\xi,z)\ra (-\xi,-z)$ and this is confirmed
by the analytical expressions both for the flavoured decay parameters determining
the wash-out and for the $C\!P$ asymmetries.

%%%%%%%%%%%%%%%%%%%%%%%%%%%%%%%%%%%%%%%%%%%%%%%%%%%%%%%%%%%%%%%%%%%%%%%%%%%%%%%%%%%%%%
\section{Leptogenesis from two RH neutrinos in models with Light Sequential Dominance}
%%%%%%%%%%%%%%%%%%%%%%%%%%%%%%%%%%%%%%%%%%%%%%%%%%%%%%%%%%%%%%%%%%%%%%%%%%%%%%%%%%%%%%

In section \ref{sec:allowed_regions} (c.f.\ Figure \ref{enlargement}) we have seen that two new favoured region for leptogenesis have appeared where
$ z \sim \pm \pi/2 $, for $\zeta = \pm 1$, and for NH. Compared to previous studies where the production of the baryon asymmetry in this region of parameters was thought to be very suppressed, we found that, due to effects from $N_2$ decays, leptogenesis is quite efficient and can be realised with comparatively low $M_1 \sim 10^{11}$ GeV. This result is particularly interesting since $ z \sim \pm \pi/2 $ corresponds to the class of neutrino mass models with Light Sequential Dominance (LSD) \cite{King:1999mb}, as we will discuss below. The dictionary between the parameter $z$ and the Sequential Dominance (SD) parameters will be given explicitly in section \ref{sec:sd_dictionary}. Finally, in \ref{sec:sd_perturbed} we will discuss the decay asymmetries in an explicit example scenario of LSD and show the enhancement of the asymmetry from $N_2$ decays analytically in terms of SD parameters and the deviation from TB mixing.

\subsection{Light Sequential Dominance}\label{sec:sd}

In models with SD, the RH neutrinos contribute to the neutrino mass matrix with ``sequential'' strength, leading to a NH. In LSD, the lightest RH neutrino $N_1$ provides the largest ``dominant'' contribution, whereas the second lightest RH neutrino contributes subdominantly. When the heaviest RH neutrino (almost) decouples, we arrive (approximately) at a 2 RH neutrino model.

To understand how SD, and in particular LSD works, we begin by writing the RH neutrino Majorana mass matrix
$M_{\mathrm{RR}}$ in a diagonal basis as

\begin{equation}\label{eq:MRR_sd}
M_{\mathrm{RR}}=
 \left(\begin{array}{ccc}
%\begin{pmatrix}
M_1 & 0 & 0 \\ 0 & M_2 & 0 \\ 0 & 0 & M_3
%\end{pmatrix},
\end{array} \right),
\end{equation}
where $M_1<M_2<M_3$. In this basis we
write the neutrino (Dirac) Yukawa matrix $\lambda_{\nu}$ in terms of $(1,3)$ column vectors $A_i,$ $B_i,$ $C_i$ as
\begin{equation}\label{eq:Ynu_sd}
Y_{\nu }= \left(\begin{array}{ccc} A & B & C
\end{array} \right)
  \label{Yukawa}
\end{equation}
in the convention where the Yukawa matrix is given in left-right convention. Explicitly we have
\begin{equation}
Y_{\nu }= \left(\begin{array}{ccc} A_1 & B_1 & C_1 \\ A_2 & B_2 & C_2 \\ A_3 & B_3 & C_3
\end{array} \right).
  \label{YukawaLSD}
\end{equation}

The Dirac neutrino mass matrix is then given by $m_{\mathrm{LR}}^{\nu}=Y_{\nu}\,v_{\mathrm{ u}}$. The term for the
light neutrino masses in the effective Lagrangian (after electroweak symmetry breaking), resulting from integrating out
the massive RH neutrinos, now reads
\begin{equation}
\mathcal{L}^\nu_{eff} = \frac{(\nu_{i}^{T} A_{i})(A^{T}_{j} \nu_{j})v^2}{M_1}+\frac{(\nu_{i}^{T} B_{i})(B^{T}_{j}
\nu_{j})v^2}{M_2} +\frac{(\nu_{i}^{T} C_{i})(C^{T}_{j} \nu_{j})v^2}{M_3} \, , \label{leff}
\end{equation}
where $\nu _{i}$ ($i=1,2,3$) are the left-handed neutrino fields. As stated above, LSD then corresponds to
$M_3\rightarrow \infty$ so that
the third term becomes negligible, with the second term subdominant and the first term dominant
\cite{King:1999mb}: \beq\label{SDcond}
\frac{A_{i}A_{j}}{M_1} \gg \frac{B_{i}B_{j}}{M_2} \gg \frac{C_{i}C_{j}}{M_3} \, .
\eeq
In addition, we shall assume that
small $\theta_{13}$ and almost maximal $\theta_{23}$ require that \beq |A_1|\ll |A_2|\approx |A_2|. \label{SD2} \eeq
Constrained Sequential Dominance (CSD) is defined as \cite{King:2005bj}:
\begin{eqnarray}
|A_{1}| &=&0,  \label{tribicondsd} \\ \text{\ }|A_{2}| &=&|A_{3}|,  \label{tribicondse} \\ |B_{1}| &=&|B_{2}|=|B_{3}|,
\label{tribicondsa} \\ A^{\dagger }B &=&0.  \label{zero}
\end{eqnarray}
CSD implies TB mixing \cite{King:2005bj} and vanishing leptogenesis if $M_3 >> M_1,M_2$ \cite{Antusch:2006cw,Choubey:2010vs}.

\subsection{An $R$ matrix dictionary for LSD}\label{sec:sd_dictionary}
According to LSD,
the ``dominant'' $N_1$, i.e.\ its mass and Yukawa couplings, governs the largest light neutrino mass $m_3$, whereas the ``subdominant'' $N_2$ governs the lighter neutrino mass $m_2$, while the decoupled $N_3$ is associated with
$m_1\rightarrow 0$.
From eq.~(\ref{R_explicit1}) it is then clear that, ignoring $m_2/m_3$ corrections,
the R-matrix for LSD takes the approximate form \cite{King:2006hn}:
\beq
R^{LSD} \approx \mathrm{diag}(\pm 1, \pm 1, 1)
\left(\begin{array}{ccc}
0 & 0 & 1 \\
0 & 1 & 0\\
1 & 0 & 0
\end{array}
\right),
\label{RLSD}
\eeq
where the four different combinations of the signs correspond physically to the four different
combinations of signs of the Dirac matrix columns associated with the lightest two RH neutrinos
of mass $M_1$ and $M_2$. The sign of the third column associated with $M_3\rightarrow \infty $ is irrelevant and
has been dropped since it
would in any case just redefine the overall sign of the Dirac mass matrix. These choices of signs are of course
irrelevant for the light neutrino phenomenology, since the effect of the orthogonal $R$ matrix cancels
in the see-saw mechanism (by definition).
The four choices of sign are also irrelevant for type I leptogenesis, since each column enters quadratically in both the
asymmetry and the washout formulas in Eqs.~(\ref{epsia}) and (\ref{Kial}),
independently of flavour or whether $N_1$ or $N_2$ is contributing.
Comparing eq.~(\ref{RLSD}) to the parameterisation of $R^{(NH)}$ for the 2 RH neutrino models in eq.~(\ref{RNH}),
we see that LSD just corresponds to $ z \sim \pm \pi/2 $ which correspond to the
new regions opened up by $N_2$ leptogenesis that were observed numerically in the previous section.
To be precise the dictionary for the
sign choices in eq.~(\ref{RLSD}) are as follows: for the $\zeta = 1$ branch,
$ z \approx \pi/2 $, corresponds to $\mathrm{diag}(1, -1, 1)$, while
$ z  \approx -\pi/2 $,  corresponds to $\mathrm{diag}(-1, 1, 1)$;
for the $\zeta = -1$ branch,
$ z  \approx \pi/2 $,  corresponds to $\mathrm{diag}(-1, -1, 1)$, while
$ z  \approx -\pi/2 $,  corresponds to $\mathrm{diag}(1, 1, 1)$.
According to the above observation, all four of these regions will contribute
identically to leptogenesis, as observed earlier in the numerical and analytical
results (i.e. giving identical results for $\zeta = \pm 1$ and $ z  \approx  \pm \pi/2$).

We may expand eq.~(\ref{RNH}) for LSD for any one of these identical regions to leading order in $m_2/m_3$.
For example consider the case $\zeta = -1$  and $ z  \approx -\pi/2 $ corresponding to
the case where all the Dirac columns have the same relative sign, $\mathrm{diag}(1, 1, 1)$.
Then expanding eq.~(\ref{RNH}) around $ z  \approx -\pi/2 $, defining $ \Delta \approx  z + \frac{\pi}{2}$,
we may write,
\beq R^{LSD} \approx
\left(\begin{array}{ccc}
0 & \Delta  & 1 \\
0 & 1 & -\Delta \\
1 & 0 & 0
\end{array}
\right).
\label{RRRR}
\eeq
Using the results in \cite{King:2010bk} we find useful analytic expressions which relate the
R-matrix angle to the Yukawa matrix elements near the CSD limit of LSD corresponding to small $\Delta $,
\bea \label{eq:Delta}
{\rm Re}(\Delta )&\approx & \frac{{\rm Re}(A^{\dagger}B)v^2}{(m_3-m_2)M_3^{1/2}M_2^{1/2}} \nonumber \\
{\rm Im}(\Delta )&\approx & \frac{{\rm Im}(A^{\dagger}B)v^2}{(m_3+m_2)M_3^{1/2}M_2^{1/2}} \,  .
\eea

Notice that $\Delta  \rightarrow 0$ when $A^{\dagger}B \rightarrow 0$ to all orders in
$m_2/m_3$. This is just the case in CSD due to eq.~(\ref{zero}).
Thus in the CSD limit of LSD eq.~(\ref{RLSD}) becomes exact \cite{King:2006hn}
to all orders in $m_2/m_3$.
Clearly, leptogenesis vanishes in CSD which can be understood from the fact that the R-matrix in CSD is real and diagonal (up to a permutation) \cite{Choubey:2010vs} or from the fact that A is orthogonal to B \cite{Antusch:2006cw}.
However in the next section we consider a perturbation of CSD,
allowing leptogenesis but preserving TB mixing.

\subsection{Example: perturbing the CSD limit of LSD}\label{sec:sd_perturbed}

Using eq.~(\ref{epsia}), we obtain, making the usual hierarchical RH neutrino mass assumption the $N_1$
contribution to the leptogenesis asymmetry parameter is given by: \be \varepsilon_{1 \alpha} \approx -\frac{3 }{16 \pi }
\frac{M_1}{M_2}\frac{1}{A^\dagger A} \mathrm{Im}\left[ A_\alpha^* (A^\dagger B)B_\alpha  \right]. \ee Clearly the
asymmetry vanishes in the case of CSD due to eq.~(\ref{zero}). In this subsection we consider
an example which violates CSD, but maintains TB mixing and stays close to LSD.

Before we turn to an explicit example, let us state the expectation for the size of the decay asymmetries.
We expect that, typically, \be \varepsilon_{1 \mu, \tau} \approx -\frac{3 }{16 \pi } \frac{m_2 M_1}{v^2}, \ \
\varepsilon_{1 e} \approx   \frac{A_1}{A_2} \varepsilon_{1 \mu, \tau}  . \label{N1} \ee The $N_2$ contribution to the
leptogenesis asymmetry parameter is given by the interference with the lighter RH neutrino in the loop via the
second term in eq.~(\ref{epsia}), which is indeed often ignored in the literature: \be \varepsilon_{2 \alpha} \approx -\frac{2 }{16
\pi } \frac{1}{B^\dagger B} \mathrm{Im}\left[ B_\alpha^* (A^\dagger B)A_\alpha  \right]. \ee This leads to typically,
\be \varepsilon_{2 \mu, \tau} \approx -\frac{1 }{16 \pi } \frac{m_3 M_1}{v^2} ,\ \ \varepsilon_{2 e} \approx
\frac{A_1}{A_2} \varepsilon_{2 \mu, \tau} \label{N2} \ee which should be compared to eq.~(\ref{N1}). The $N_2$ contribution to the decay asymmetries looks larger than the $N_1$ contribution.

To compare the two asymmetries and the produced baryon asymmetry explicitly, let us now calculate the final asymmetries in a specific perturbation of the Light CSD form. As an example, we may consider
\begin{eqnarray}
(A_1,A_2,A_3) &=& (0,a,a) \\
(B_1,B_2,B_3) &=& (b,b+q,-b+q)
\end{eqnarray}
such that
\begin{equation}
Y_{\nu }= \left(\begin{array}{ccc} 0 & b & C_1 \\ a & b+q & C_2 \\ a & -b+q & C_3
\end{array} \right).
  \label{YukawaLSD_perturbed}
\end{equation}
Providing $|q|\ll |b|$,
this perturbation of CSD but stays close to LSD and allows non-zero leptogenesis.
Interestingly this perturbation of CSD also preserves TB mixing as discussed in \cite{King:2010bk},
where more details can be found.
Note that $z$ is given by eq.~(\ref{eq:Delta}) and therefore depends on $a,b$ and $q$.

For our example, we now obtain (assuming real $a$ and neglecting $q$ in $B^\dagger B$):
\be
\varepsilon_{1 \gamma} \approx -\frac{3 }{16 \pi } \frac{m_2 M_1}{v^2} \frac{\mathrm{Im}[q \, b + q^2]}{B^\dagger B}, \ \
\varepsilon_{1 \tau} \approx  -\frac{3 }{16 \pi } \frac{m_2 M_1}{v^2} \frac{\mathrm{Im}[- q \, b + q^2]}{B^\dagger B}   .
\ee
The $N_2$ contribution to the
leptogenesis asymmetry parameter is given by the interference with the lighter RH neutrino in the loop via the
second term in eq.~(\ref{epsia}):
\be
\varepsilon_{2 \gamma} \approx -\frac{2 }{16 \pi } \frac{m_3 M_1}{v^2} \frac{\mathrm{Im}[q \, b^*]}{B^\dagger B} ,\ \
\varepsilon_{2 \tau} \approx - \varepsilon_{2 \gamma}
\ee
For the washout parameters, we obtain:
\begin{eqnarray}
K_{1 \gamma} = K_{1 \tau} \approx \frac{m_3}{m_*}
\end{eqnarray}
and
\begin{eqnarray}
K_{2 \gamma} \sim K_{2 \tau} \approx \frac{m_2}{m_*}\:.
\end{eqnarray}
The parameter $p_{12}$ is given by (neglecting $q$ in the last step)
\begin{eqnarray}
p_{12} \approx
-\frac{|A_1 B_2 + A_2 B_1 |^2}{(|A_1|^2 + |A_2|^2) (|B_1|^2 + |B_2|^2)} \approx \frac{1}{2}
\end{eqnarray}
For the final asymmetries from $N_1$ decay this means
\begin{eqnarray}
N_{B-L}^{f(1)} \approx -(\varepsilon_{1\gamma} + \varepsilon_{1\tau})\,
\kappa_{1 \gamma} \approx
2 \frac{3 }{16 \pi } \frac{m_2 M_1}{v^2} \frac{\mathrm{Im}[q^2]}{B^\dagger B} \kappa(m_3/m^*)
\end{eqnarray}
whereas
\begin{eqnarray}
N_{B-L}^{f(2)} \approx -(1-p_{12}) \varepsilon_{2\gamma}\,\kappa_{2 \gamma} 
\approx
\frac{1 }{2 } \frac{2 }{16 \pi } \frac{m_3 M_1}{v^2} \frac{\mathrm{Im}[q \, b^*]}{B^\dagger B}
\kappa(m_2/m^*) \:.
\end{eqnarray}
So we can estimate:
\begin{eqnarray}
\frac{N_{B-L}^{f(2)}}{N_{B-L}^{f(1)}} \approx  
 \frac{m_3}{m_2} \frac{\kappa(m_2/m^*)}{\kappa(m_3/m^*)}  \frac{\mathrm{Im}[q \, b^*]}{\mathrm{Im}[6 \, q^2]}.
\end{eqnarray}
We see that, as already anticipated in the beginning of this subsection, there is an enhancement of the asymmetry from the $N_2$ decay by a factor of $\frac{m_3}{m_2}$ (from the decay asymmetries). Furthermore, there is another enhancement factor from the efficiency factor $\kappa$ given by $\frac{\kappa(m_2/m^*)}{\kappa(m_3/m^*)} $. Both terms imply an enhancement of a factor of 5 each. 
Finally, the factor $\frac{\mathrm{Im}[q \, b^*]}{\mathrm{Im}[6 \, q^2]}$ can get large for small $q$, i.e.\ close to the CSD case. However, of course, closer to the CSD case the decay asymmetries get more and more suppressed.

In summary, in models with Light Sequential Dominance (LSD) the asymmetry from the $N_2$ decays is generically larger than the asymmetry from $N_1$ decays, in agreement with the results obtained in the previous sections in the R matrix parameterisation. We like to emphasise that in order to calculate the prospects for leptogenesis in models with LSD (in the two flavour regime), it is thus crucial to include the $N_2$ decays (which have previously been neglected).

%%%%%%%%%%%%%%%%%%%%%
\section{Conclusions}
%%%%%%%%%%%%%%%%%%%%%
We have revisited leptogenesis in the minimal non-supersymmetric type I see-saw mechanism with
two hierarchical RH neutrinos  ($M_2\gtrsim 3\,M_1$), including flavour effects and allowing
both RH neutrinos $N_1$ and $N_2$ to contribute, rather than just the lightest RH
neutrino $N_1$ that has hitherto been considered.

We emphasise two crucial ingredients of our analysis:
i) the flavoured $C\!P$ asymmetries have been
calculated taking into account also terms that cancel in the total $C\!P$ asymmetries
\cite{dirac} and that have been so far neglected within the two RH neutrino model;
ii) Part of the asymmetry produced from $N_2$ decays, that one orthogonal
in flavour space to the lepton flavour $\ell_1$ produced by $N_1$ decays, escapes the
$N_1$ washout \cite{N2importance}.

Defining four benchmark points corresponding to a range of PMNS parameters,
we have performed scans over the single
complex angle $z$ of the orthogonal matrix $R$ for each of the two physically distinct branches
$\zeta = \pm 1$. For the case of a normal mass hierarchy,
for each benchmark point we found that in regions around $z\sim \pm \pi /2$, the
$N_2$ contribution can dominate the contribution to leptogenesis.
For benchmark B corresponding to a large reactor angle and zero low energy CP violation
we found that the lightest RH
neutrino mass may be decreased by about an order of magnitude in these regions,
down to $M_1\sim 1.3 \times 10^{11}\,{\rm GeV}$ for vanishing initial $N_2$-abundance,
with the numerical results supported by analytic estimates. Other benchmarks with
smaller reactor angle and/or low energy CP violating phases switched on exhibit similar results.

These $N_2$-dominated regions around $z\sim \pm \pi /2$ are quite interesting
since they correspond to light sequential dominance in the hierarchical limit where the
atmospheric neutrino mass $m_3$ arises dominantly from the lightest RH neutrino
of mass $M_1$, the solar neutrino mass $m_2$ arises dominantly from the second
lightest RH neutrino of mass $M_2$, and the lightest neutrino mass of $m_1$
is negligible due to a very large RH neutrino of mass $M_3$.
Such a scenario commonly arises in unified models based on a natural application
of the see-saw mechanism \cite{Antusch:2006cw,Antusch:2004gf}
so the new results in this paper may be relevant to
unified model building in large classes of models involving a NH.

\subsection*{Acknowledgments}

PDB acknowledges financial support from the NExT Institute and SEPnet.
SA acknowledges partial support by the DFG cluster of excellence `Origin and Structure of the Universe'.
DAJ is thankful to the STFC for providing studentship funding. PDB and SFK were partially supported by the STFC Rolling Grant
ST/G000557/1 and SFK was partially supported by the EU ITN grant UNILHC 237920 (`Unification in the LHC era').

\end{document}